\documentclass[11pt, a4paper, logo, copyright]{googledeepmind}

\pdftrailerid{redacted}

\makeatletter
\renewcommand\bibentry[1]{\nocite{#1}{\frenchspacing\@nameuse{BR@r@#1\@extra@b@citeb}}}
\makeatother

\usepackage{kantlipsum, lipsum}
\usepackage{dsfont}
\usepackage{gdm-colors}
\usepackage[utf8]{inputenc}   
\usepackage{newunicodechar}   
\usepackage{wasysym,marvosym}
\usepackage{ulem}
\usepackage{hyperref}       
\usepackage{algorithmicx}
\usepackage{algpseudocode}
\usepackage{multirow}
\usepackage{geometry} 
\usepackage[rightcaption]{sidecap} 

\usepackage{tikz}
\usetikzlibrary{shapes.geometric, arrows.meta, positioning}
\usepackage{amsmath}
\usepackage{algorithm}
\usepackage{algpseudocode}
\usepackage{listings}
\usepackage{enumitem} 
\usepackage{lipsum}
\usepackage[most]{tcolorbox}
\definecolor{thinkcolor}{RGB}{227,196,144}
\definecolor{observecolor}{RGB}{153,201,227}
\definecolor{explorecolor}{RGB}{178,217,200}

\newcounter{caseexample}[section]

\newcounter{promptexample}[section]

\usepackage{array, multirow, tabularx, booktabs, makecell}

\newcommand{\assignmentQuestionName}{Question} 

\usepackage{booktabs}
\usepackage{arydshln}
\usepackage{dashrule}
\usepackage{twemojis}

\usepackage[authoryear, sort&compress, round]{natbib}

\usepackage{bbding}
\usepackage[T1]{fontenc}    
\usepackage{url}            
\usepackage{booktabs}       
\usepackage{nicefrac}       
\usepackage{microtype}      
\usepackage{amsmath}
\usepackage{graphicx}
\usepackage{multicol}
\usepackage[nameinlink]{cleveref}
\usepackage{bbm}
\usepackage{multirow}
\usepackage{soul}
\usepackage{float}
\usepackage{wrapfig}
\usepackage{blindtext}
\usepackage{tablefootnote}
\usepackage{amsfonts}
\usepackage[flushleft]{threeparttable}
\usepackage{colortbl}
\usepackage{mathtools}
\usepackage{bm}
\usepackage{CJKutf8}
\usepackage{makecell}
\usepackage{caption}
\usepackage{capt-of}
\usepackage{array}
\usepackage{calc}      
\usepackage{caption}   
\usepackage{subcaption}  
\usepackage[bottom]{footmisc}
\usepackage{fontawesome}
\usepackage{tabularx}        
\usepackage{siunitx}

\definecolor{lav}{HTML}{8E63C1}  
\definecolor{blu}{HTML}{6DA8DA}  
\definecolor{ros}{HTML}{CE1F49} 
\definecolor{split}{HTML}{E0FFE0}
\definecolor{numeric}{HTML}{E0E0FF}
\definecolor{random}{HTML}{FFF0E0}
\definecolor{hash}{HTML}{E0FFE0}
\definecolor{leaf}{HTML}{E0F0F0}
\definecolor{wire}{HTML}{E8E8E8}
\definecolor{stable}{HTML}{FFD1D1}
\definecolor{hlcolor}{RGB}{206,32,74}

\newcolumntype{C}{>{\centering\arraybackslash}X}
\newcolumntype{L}{>{\raggedright\arraybackslash}X}

\newcommand{\pilot}[1][]{\textbf{\texttt{PILOT#1}}}
\newcommand{\oldway}[1]{\textbf{\textcolor{teal!85!black}{#1}}}
\newcommand{\newway}[1]{\textbf{\textcolor{orange!85!black}{#1}}}
\newcommand{\rolestar}{\textcolor{orange!85!black}{$\bigstar$}}
\newcommand{\hlnum}[1]{\textbf{\textcolor{hlcolor}{#1}}}
\newcommand{\virtue}[1]{\textbf{\textit{#1}}}

\newtcbox{\pill}[1][purple]{on line, nobeforeafter,
  colback=#1!16, colframe=#1!16, boxrule=0pt, arc=4.5pt,
  left=1pt, right=1pt, top=1pt, bottom=1pt,
  fontupper=\small\ttfamily\bfseries, coltext=#1!60!black}

\newtcbox{\nill}[1][purple]{on line, nobeforeafter,
  colback=#1!16, colframe=#1!16, boxrule=0pt, arc=4.5pt,
  left=0.5pt, right=0.5pt, top=0pt, bottom=0pt,
  fontupper=\small\ttfamily\bfseries, coltext=#1!60!black}

\title{PILOT Technical Report}

\author{PILOT Team}

\begin{abstract}
Existing agentic approaches for recommendation system optimization remain fundamentally reactive—they adjust parameters in response to observed metric changes but lack the ability to proactively design controlled experiments, personalize strategies at the user-segment level, or accumulate reusable experimental methodology across tasks.
We present \pilot{} (\textbf{P}roactive \textbf{I}nsight \textbf{L}earner for \textbf{O}nline \textbf{T}ree-Experiments), an LLM-agent framework that organizes three roles within a constrained control loop where deterministic services enforce all safety, statistical, and permission boundaries: (1)~an \newway{Experiment Manager} that drives the full experiment lifecycle---task intake, observation governance, anomaly recovery, and postmortem---by selecting only from a rule-generated legal-command envelope; (2)~a \newway{Search Planner} that proposes candidate decision trees for user-segment-level personalization, invoked only when the Manager requests planning; and (3)~a \newway{Memory Curator} that asynchronously distills experiment outcomes into strategy-level domain knowledge and provenance-tracked methodology, failure-isolated from the main loop. The Manager makes the agent proactive, the Planner enables population-level personalization beyond global tuning, and the Curator turns every completed task into a learning opportunity for the next.
Deployed on Taobao's platform with 5 experimental buckets, \pilot{} is compared against ROAM (\textbf{R}eactive \textbf{O}ptimization with \textbf{A}gent-driven \textbf{M}oves), a free-exploration agent without lifecycle governance or structured hypothesis testing. \pilot{} achieves up to \hlnum{+1.40\%} IPV, \hlnum{+1.60\%} Core IPV, \hlnum{+0.96\%} transaction count, and \hlnum{+1.50\%} transaction amount, improving over ROAM's best results (\hlnum{+1.00\%} IPV, \hlnum{+0.90\%} Core IPV, \hlnum{+0.60\%} transaction count, \hlnum{+1.13\%} transaction amount) while raising search efficiency from \hlnum{53.3\%} to \hlnum{93.3\%} (\hlnum{+40pp}), with no human intervention throughout the experimental cycle.
\end{abstract}

\begin{document}
\begin{CJK*}{UTF8}{gkai}

\hspace{0.5cm}\maketitle
\vspace{-13.2cm}

\hspace{1.75cm}\includegraphics[height=1.5cm]{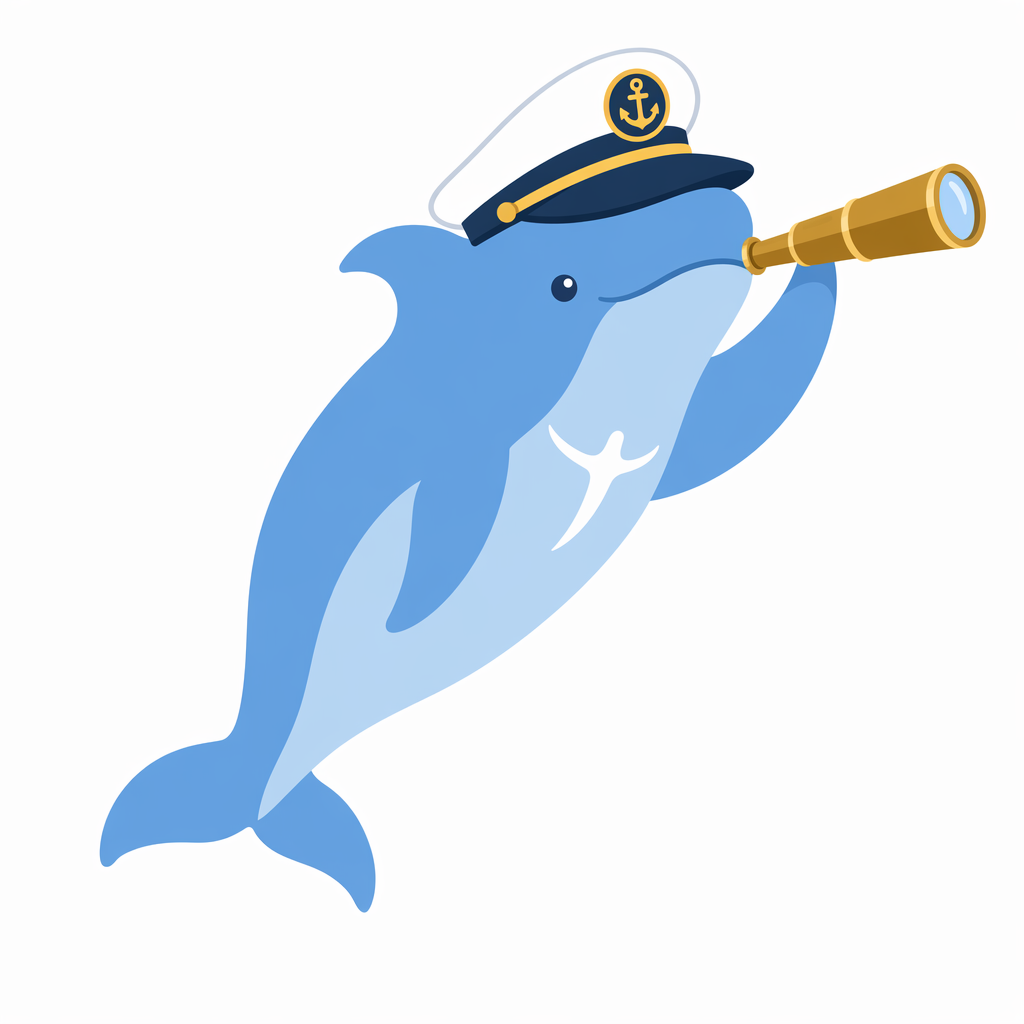}
\vspace{12cm}

\begin{figure}[htbp]
    \centering
    \includegraphics[width=\linewidth]{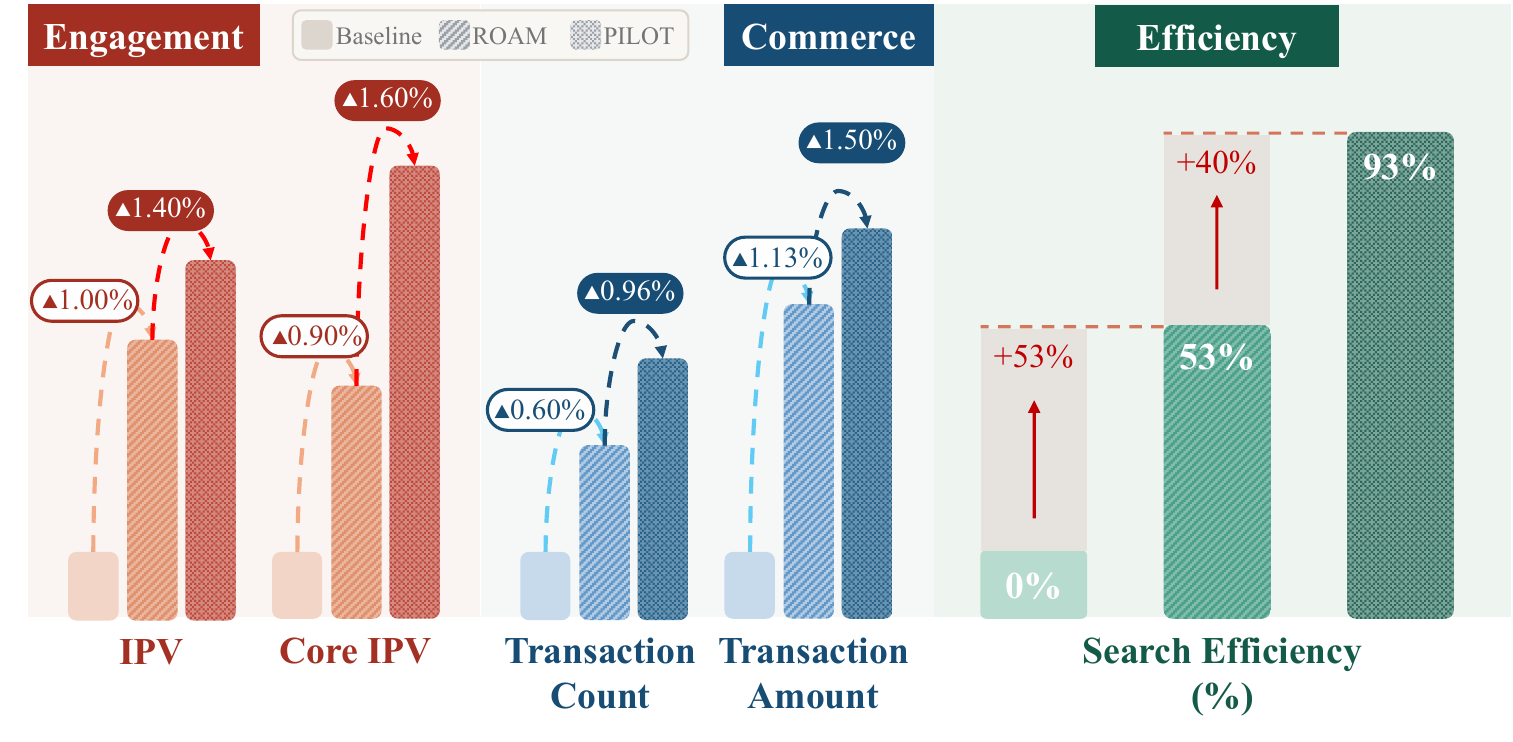}
    \caption{Online A/B results for \pilot{} and ROAM on Taobao Homepage Guess You Like.}
    \label{fig:dream_teaser}
\end{figure}

\newpage
\setcounter{tocdepth}{2}

\tableofcontents 

\newpage

\section{Introduction}
\label{sec:introduction}

Industrial recommendation platforms run online experimentation as a continuous control loop: buckets are split, strategies are deployed, and metrics are watched to decide the next adjustment. This loop has traditionally been driven by human experimenters, and as the number of concurrent experiments and user segments grows, manual design and iteration no longer scale.

Agentic approaches for recommendation have started to move such optimization decisions from humans to LLM agents. Beyond agents that select items with memory and tool use~\citep{memrec2026,recnet2026,steam2026,amem4rec2026,sager2026,chainrec2026,recthinker2026,rrcm2026,twistar2026,reasonrec2026}, simulate users for training and evaluation~\citep{agentcf2023,agentictagger2026,agentgr2026,alignuser2026,abagent2026}, or converse with users~\citep{idss2026,recoworld2025,iagent2025,recbot2025}, a growing line of work treats the agent as a coordinator that acts on the recommendation system itself rather than on the recommendation list~\citep{selfevolvingrecsys2026,agenticrectune2026,nova2026,automodel2026,evorec2026,agentx2026,sortify2026}. This coordinator setting is closest to ours, yet it shares three key limitations that constrain further progress:

\begin{itemize}
    \item[\textcolor{teal!85!black}{\textbf{L1}}] \oldway{Reactive response.} Existing coordinators act as fast responders to observed metric change or detected failure, not as experimenters that propose and validate a hypothesis before it is deployed.
    \item[\textcolor{teal!85!black}{\textbf{L2}}] \oldway{Global-only strategies.} Optimization is applied at the level of the whole system or bucket, with no mechanism for the agent to personalize strategies to sub-populations of users within a bucket.
    \item[\textcolor{teal!85!black}{\textbf{L3}}] \oldway{No methodology accumulation.} Memory and self-evolving skills are increasingly common, but what accumulates is knowledge about the content of a single task; this knowledge is not distilled into the agent's own experimental methodology, so it does not compound into a reusable capability that makes the agent faster or better at running the next experiment.
\end{itemize}

To address these limitations, we present \pilot{} (\textbf{P}roactive \textbf{I}nsight \textbf{L}earner for \textbf{O}nline \textbf{T}ree-Experiments), an LLM-agent framework that operates like a human experimenter rather than \oldway{a metric responder}: it \newway{manages experiment lifecycles}, \newway{proposes population-level personalization via decision trees}, and \newway{distills reusable knowledge from each experiment}. \pilot{} organizes three LLM roles within a constrained control loop, where deterministic services enforce all safety, statistical, and permission boundaries.

The architecture of \pilot{} embodies the following three key capabilities—one for each LLM role—that directly answer \textbf{L1}--\textbf{L3} above:

\begin{itemize}
    \item[\rolestar] \newway{Experiment Manager (\S\ref{sec:manager}).} The Experiment Manager drives the full experiment lifecycle --- task intake, playbook assembly, observation governance, anomaly recovery, scale validation, and postmortem --- rather than responding to metric alerts after the fact. At each step, a deterministic Manager Guard \newway{generates a legal-command envelope from the current state and frozen policies}, and the Manager selects only from this envelope (\virtue{bounded}). High-risk decisions --- estimand, guardrails, budget, final confirmation --- require explicit user input rather than silent inference (\virtue{safe}), and a fixed authority hierarchy ensures that frozen safety policies and user-approved specifications always override agent reasoning (\virtue{auditable}).
    \item[\rolestar] \newway{Search Planner (\S\ref{sec:planner}).} Global-only strategies apply a single strategy to an entire bucket, even though users within it can need different treatment. The Search Planner proposes candidate decision trees that \newway{personalize strategies at the user-segment level}, binding a distinct strategy bundle to each leaf so differentiation only happens where user behavior actually differs (\virtue{personalized}). Candidates are drawn from three lanes --- a deterministic baseline, LLM-proposed hypotheses with evidence references, and a bounded exploration slot --- and every candidate must clear a deterministic validator before entering the experiment (\virtue{validated}). The Planner is invoked only when the Experiment Manager explicitly requests planning; it cannot manage lifecycle, adjust observation windows, or promote Champions (\virtue{scoped}).
    \item[\rolestar] \newway{Memory Curator (\S\ref{sec:curator}).} The Memory Curator operates asynchronously after evidence settlement, \newway{distilling experiment outcomes into two distinct stores}: a strategy evidence store that records bundle-contrast effects with population scope and evidence-role tagging (\virtue{contrastive}), and a methodology experience store that tracks how to run experiments with source-type provenance and a governed confidence lifecycle from draft to supported to approved (\virtue{governed}). Each task forks an isolated branch from the shared knowledge base and merges back only validated entries at task end. The next task therefore starts from a richer base without cross-task contamination (\virtue{compounding}).
\end{itemize}

\pilot{} has been deployed on Taobao's recommendation platform across 5 experimental buckets, compared against ROAM (\textbf{R}eactive \textbf{O}ptimization with \textbf{A}gent-driven \textbf{M}oves), an agent-based configuration that performs free exploration without lifecycle governance or structured hypothesis testing. \pilot{} improves the best-bucket results from \hlnum{+1.00\%} to \hlnum{+1.40\%} IPV, from \hlnum{+0.90\%} to \hlnum{+1.60\%} Core IPV, from \hlnum{+0.60\%} to \hlnum{+0.96\%} transaction count, and from \hlnum{+1.13\%} to \hlnum{+1.50\%} transaction amount, while raising search efficiency from \hlnum{53.3\%} to \hlnum{93.3\%} (\hlnum{+40}pp), requiring no human intervention throughout the experimental cycle.

The remainder of this report is organized as follows. \S\ref{sec:framework} overviews the \pilot{} framework and its constrained control loop. \S\ref{sec:manager}--\S\ref{sec:curator} detail the Experiment Manager, Search Planner, and Memory Curator respectively. \S\ref{sec:experiments} reports online A/B results, ablations, and a case study of hypothesis evolution, and \S\ref{sec:conclusion} concludes.
\section{The PILOT Framework}
\label{sec:framework}

\begin{figure*}[t]
\centering
\includegraphics[width=0.95\linewidth]{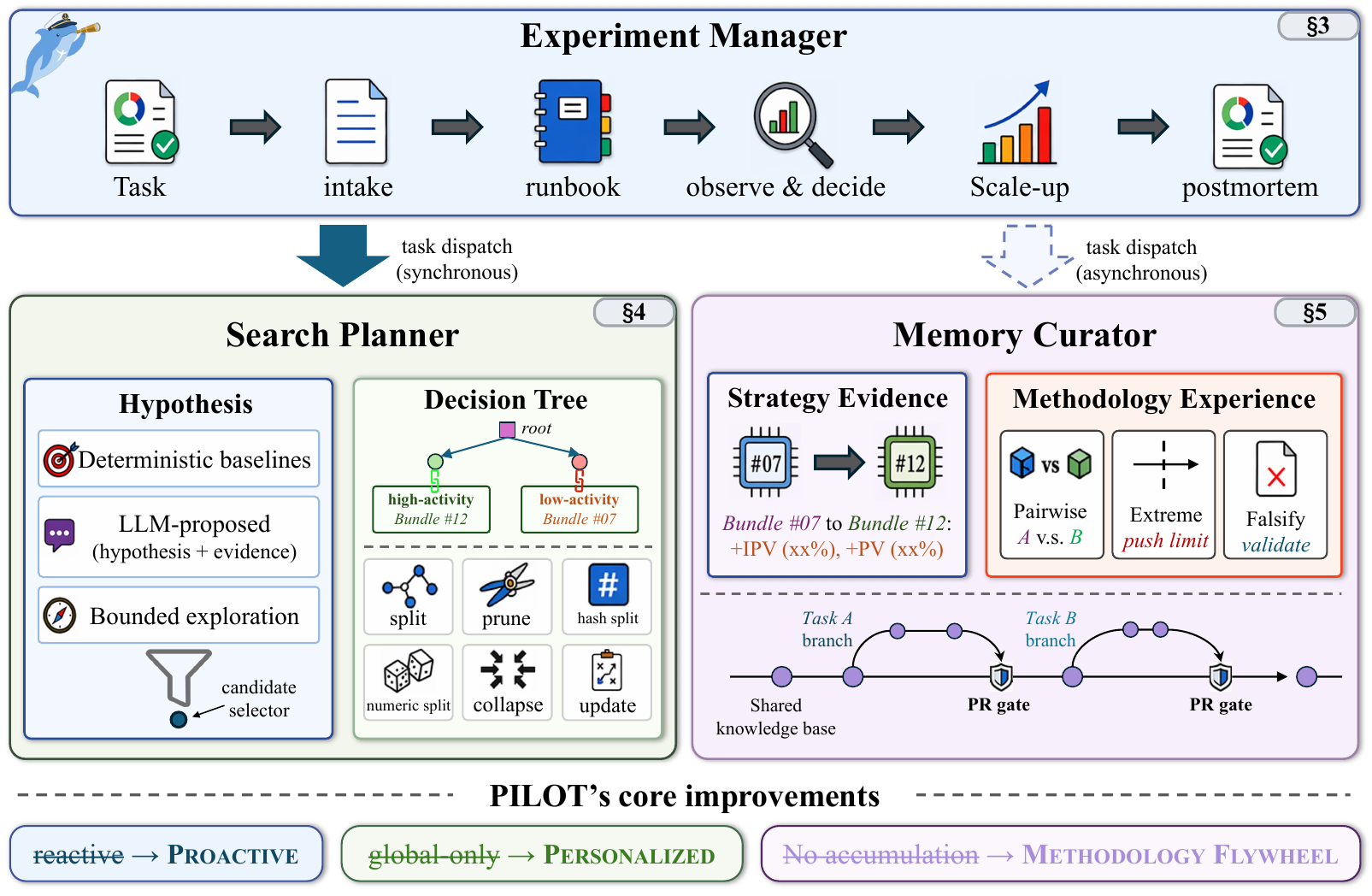}
\caption{The framework overview of \pilot{}. \pilot{}'s three LLM roles within the constrained control loop, each addressing one limitation of existing agentic coordinators: the Experiment Manager drives proactive lifecycle control, the Search Planner enables population-level personalization via decision trees, and the Memory Curator lets experimental methodology evolve across tasks.}
\label{fig:pilot_framework}
\end{figure*}

\pilot{} places the LLM at three positions that most need judgment: lifecycle management, candidate proposal, and evidence distillation, and wires all three into one \textbf{constrained control loop} (Figure~\ref{fig:pilot_framework}). Everything else in the system is deterministic.

\label{sec:framework_control}

An event (a user request, a platform signal, a timer, or an executor callback) triggers the Manager Guard to work out which actions are currently allowed: a short list of legal next steps, derived purely from the state machine and frozen policy, without LLM involvement. The Experiment Manager then picks one action from that list, optionally delegating to the Search Planner. The Guard checks the pick against the current state, its parameter bounds, and any outstanding certificates; the State Committer then applies it atomically and is the only component allowed to write state.

Six deterministic services carry out this loop: the Manager Guard and State Committer described above, a Statistics Engine that produces immutable decision certificates, a Contract Builder that writes down the pre-registered rules for judging an experiment's outcome, a validator that checks any candidate the Search Planner proposes, and a builder that enumerates which actions are currently legal. \textbf{LLM roles only choose among options these services already generated}: they cannot invent a command, alter a statistical conclusion, or step outside a permission boundary. Beneath these services, a foundation layer (Figure~\ref{fig:pilot_framework}, bottom) provides persistent storage---a Memory store for fork-merge knowledge, a StrategyHub for bundle definitions and templates, and a Knowledge Base for domain knowledge---along with an Online Runtime that handles population routing and bundle dispatch at serving time.

\subsection{Experiment Manager}
\label{sec:framework_manager}

The Experiment Manager is the semantic owner of the entire experiment lifecycle: task intake, assembling an operating playbook for the task, progressing the experiment by choosing only from the actions currently allowed, observation governance, anomaly recovery, scale validation, and postmortem. It shifts the agent from reactive parameter tuning to proactive experiment management.

Task intake ends with a frozen task specification agreed with the user, and an operating playbook assembled from the user's own instructions together with methodology that has proven reliable on past tasks. Lifecycle progression then moves through four time layers that are easy to conflate but distinct in meaning: one look is a single data read paired with a state judgment, one wave is a batch of co-launched \pill[lav]{Challengers}, one epoch is a window in which the \pill[lav]{Champion}, its estimator, and the decision policy are frozen so every \pill[lav]{Challenger} launched inside it shares one common comparison, and one search iteration is a complete plan-launch-observe-decide cycle. Observation governance tracks every look's outcome as provisional until a result is judged final; only a final result may update Memory or unlock the next round of planning. \S\ref{sec:manager} details the full lifecycle.

\subsection{Search Planner}
\label{sec:framework_planner}

The Search Planner is the candidate proposer: it reads the current \pill[lav]{Champion}, the list of currently legal actions, evidence, and Memory, and returns a shortlist of legal candidates together with hypotheses and evidence references. It is invoked only when the Experiment Manager explicitly asks for a plan, and it cannot manage lifecycle, adjust observation windows, allocate traffic, or promote Champions.

Every candidate it proposes is a \pill[lav]{Challenger}: a variant that carries exactly one atomic action relative to the current \pill[lav]{Champion}, the best confirmed tree and the platform's search reference, itself compared against a fixed baseline bucket, \pill[lav]{B0}, that anchors the platform and stays invariant across rounds. Before a \pill[lav]{Challenger} is deployed, a set of decision rules is agreed in advance and bound to it, fixing what counts as a positive, negative, inconclusive, or extended result, so the rules cannot be adjusted once the data starts coming in. \S\ref{sec:planner} details the decision-tree model, the space of legal actions, and three-lane candidate generation.

\subsection{Memory Curator}
\label{sec:framework_curator}

The Memory Curator is the asynchronous evidence distiller: after evidence settlement, it organizes experiment outcomes into two memory stores, one holding bundle-contrast domain knowledge with evidence-role tagging, and the other holding provenance-tracked experimental methodology with a governed confidence lifecycle. It is failure-isolated: a Memory Curator failure degrades knowledge accumulation, not the experiment itself.

The first store never records a strategy's effect on its own; every entry contrasts one strategy bundle against another and tags whether the underlying evidence supports a causal claim or only a diagnostic one. The second store tracks where a piece of methodology came from and whether it has only been tried once or has proven reliable across tasks, and only proven methodology is allowed to feed future playbooks without a human in the loop. \S\ref{sec:curator} details the two memory stores, fork-merge isolation, and write governance.

\section{Experiment Manager}
\label{sec:manager}

This section details how the Experiment Manager hosts one complete experiment: converging a task into a frozen specification and playbook (\S\ref{sec:manager_intake}), observing each round and deciding when to advance it (\S\ref{sec:manager_lifecycle}), and confirming and delivering the result once a round reaches its target (\S\ref{sec:manager_evidence}).

\subsection{Experiment Preparation}
\label{sec:manager_intake}

Before any search activity begins, the Experiment Manager turns a raw request into three frozen artifacts: it first understands what the user wants, then assembles how the task should be run, and finally fixes the order in which these artifacts can override one another.

\paragraph{Task specification.} The Experiment Manager is first a dialogue agent. A user's objective arrives as unstructured natural language, and the Manager cannot forward it to the Search Planner directly: it must converge the dialogue into a versioned task specification, denoted $\mathcal{T}$, before any search activity begins. A task specification $\mathcal{T}$ is a versioned record covering the fields listed in Table~\ref{tab:task_spec}. Once frozen, $\mathcal{T}$ is immutable for the remainder of the task, and no running experiment wave reads an unfrozen draft of it. Convergence proceeds through four completion states, \pill[ros]{incomplete}, \pill[ros]{awaiting}, \pill[blu]{ready}, and \pill[ros]{frozen}, each requiring the missing or ambiguous fields identified at the previous state to be resolved before advancing. For a subset of high-risk fields, the Manager must not silently infer a value: it must issue a structured request that lists the allowed values, states a recommended default, and sets a response deadline, and $\mathcal{T}$ cannot advance past \pill[ros]{awaiting} until the user responds.

\begin{table}[t]
\centering
\caption{Fields of a task specification $\mathcal{T}$, grouped by category and confirmation tier.}
\label{tab:task_spec}
\small
\begin{tabularx}{\textwidth}{l X l}
\toprule
\textbf{Category} & \textbf{Fields} & \textbf{Tier} \\
\midrule
\rowcolor{gray!6}
Population and search & Fixed baseline (\pill[lav]{B0}), initial \pill[lav]{Champion}, search universe & Standard \\
\addlinespace
\rowcolor{gray!6}
Governance & Scale validation goal and policy, notification policy, user methodology references & Standard \\
\addlinespace
\rowcolor{teal!10}
Budget and schedule & Observation schedule constraints, start and deadline, and traffic, bucket, and round budget & Mixed \\
\addlinespace
\rowcolor{teal!10}
Objective and metrics & Objective, minimum business effect, and estimand and primary metric & Mixed \\
\addlinespace
\rowcolor{orange!10}
Population and search & Randomization and exposure spec & High-risk \\
\addlinespace
\rowcolor{orange!10}
Governance & Guardrail and stop policy, approval matrix & High-risk \\
\bottomrule
\end{tabularx}
\end{table}

\paragraph{Playbook assembly.} Converging the dialogue into $\mathcal{T}$ tells the Manager what the task should test, but not how to run it. Before the task starts, the Manager also prepares the operating space the task will run in: \textbf{it selectively pulls together the business knowledge, memory, analysis methods, strategies, and methodology relevant to this task}, drawing first on whatever the user has already confirmed in the current dialogue, and filling any remaining gaps with what has proven reliable on past tasks. How this pool of methodology is sourced, governed, and kept trustworthy across tasks is the subject of \S\ref{sec:curator}; here it is enough to know that the outcome of this preparation is the playbook $\mathcal{P}$, the operating manual the task consumes at runtime. Before $\mathcal{P}$ is frozen, the Manager Guard checks that it covers every decision the task will face and that nothing in it exceeds the permissions of its source.

\paragraph{Authority hierarchy.} Once $\mathcal{T}$ and $\mathcal{P}$ are frozen, the Manager's choices for the remainder of the task follow a fixed order of precedence: the frozen safety and statistical policy always overrides the frozen task specification, which overrides the frozen playbook, which overrides approved methodology, which in turn overrides a provisional or unconfirmed hint, which overrides the Agent's own reasoning. A source higher in this order can never be overridden by one lower in it, even when the Agent's own reasoning points elsewhere.

\subsection{Observation and Advancement}
\label{sec:manager_lifecycle}

\begin{figure*}[h]
    \centering
    \includegraphics[width=0.9\linewidth]{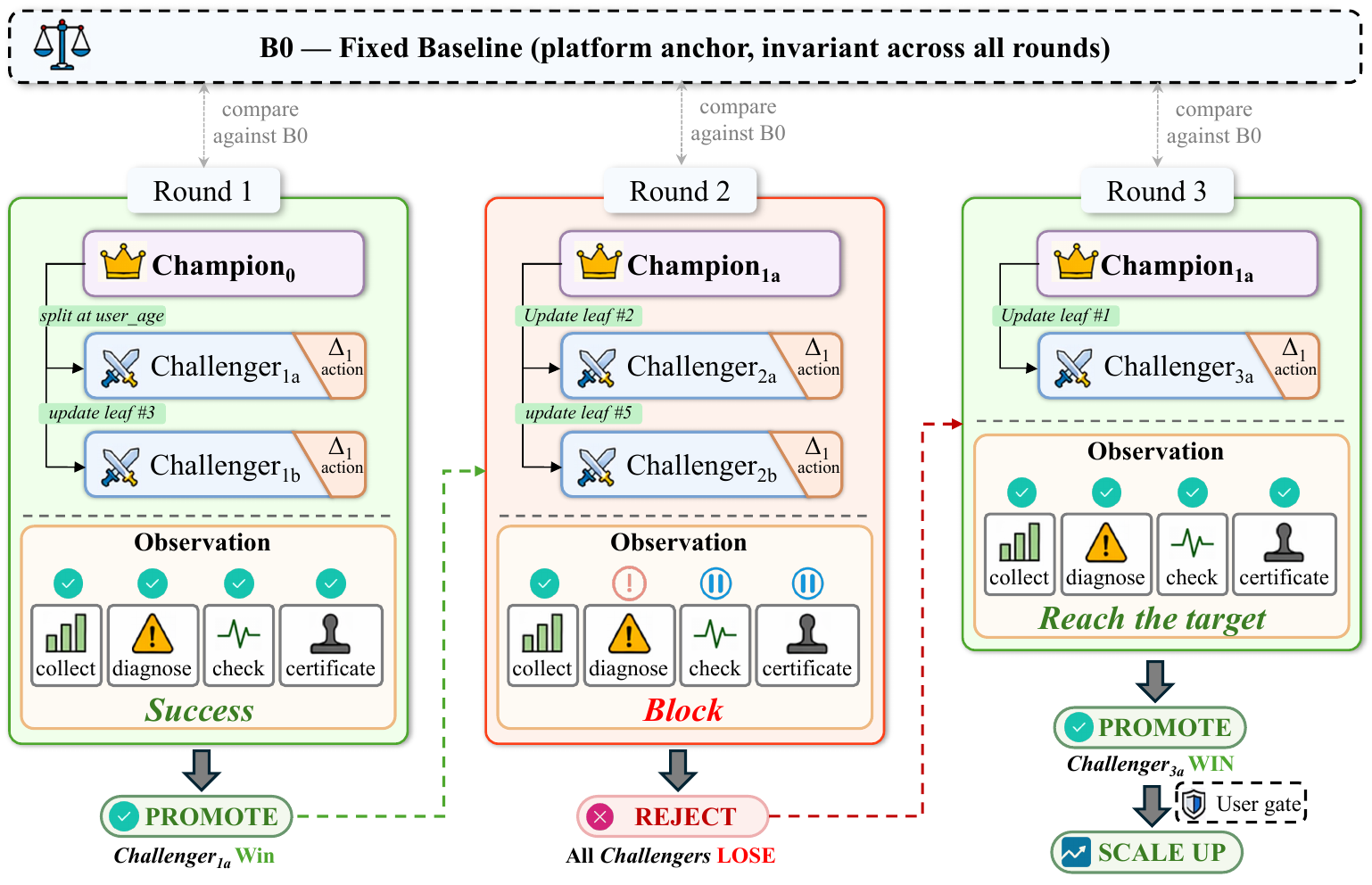}
    \caption{Champion--Challenger round lifecycle. Each round pits the current \pill[lav]{Champion} against one or more \pill[lav]{Challengers} under a fixed observation window, ending in a \pill[blu]{promote}, \pill[ros]{reject}, or \pill[blu]{continue} certificate. A promoted \pill[lav]{Challenger} becomes the next round's \pill[lav]{Champion}; a rejected one leaves the \pill[lav]{Champion} unchanged. The fixed baseline \pill[lav]{B0} anchors all comparisons across rounds.}
    \label{fig:round_lifecycle}
\end{figure*}

Figure~\ref{fig:round_lifecycle} illustrates the full round lifecycle across three example rounds. Every \pill[lav]{Challenger} is derived from the current \pill[lav]{Champion} by \textbf{applying exactly one atomic action (e.g., split on user\_age, update\_strategy leaf~\#3)}, denoted by the $\Delta$1 badge on each Challenger box. Within each round, the observation panel runs a four-step pipeline---collect metrics, diagnose anomalies, check data health, certificate decision---before the Manager may issue a certificate. As the figure shows, this pipeline is not always uniform: Round~1 closes cleanly, Round~2 detects a sample-ratio mismatch that extends the observation window before the round can close, and Round~3 closes cleanly after an extended window. When a round ends with a \pill[blu]{promote} certificate, the winning Challenger ascends to become the next round's \pill[lav]{Champion} (as in Rounds~1 and~3); when it ends with a \pill[ros]{reject}, the \pill[lav]{Champion} carries over unchanged (as in Round~2; note that the subscript in the figure tracks the round count, not whether the underlying tree changed). After the final promotion, a "Target Met?" gate checks whether the task objective in $\mathcal{T}$ is satisfied: if yes, the task exits the search loop into staged scale-up, independent confirmation on $D_{\text{confirm}}$, and production delivery; if no and budget remains, the dashed loop-back arrow returns the task to a new round. If no budget remains, the search halts.

\paragraph{Observing a round.} Once a round is running, the Experiment Manager watches it on a fixed observation plan: at each look it collects the round's metrics, diagnoses any fluctuation or assignment anomaly (a sample-ratio mismatch, a data delay, an exposure break), and decides the round's window, whether to keep waiting, extend, or close. These calls rest only on data-health signals such as freshness, maturity, sample size, and integrity, never on the treatment lift direction, which stays hidden until a close is actually warranted; an ``almost significant'' reading, a favorable date cherry-picked after the fact, or a window trimmed to flatter the result are all out of bounds. When a look shows an anomaly that looks like noise, it is marked diagnostic-only, so it does not count as evidence, and the window is extended so a clean look can still come in; a recurring anomaly, or one that breaches a hard bound, pauses the round without closing it and escalates to a human. An unresolved anomaly never gets to influence whether the \pill[lav]{Champion} changes.

\paragraph{Round certificates.} When the signals clear, the Manager closes the round with one of three certificates. A \pill[blu]{continue} certificate keeps the round running and schedules the next look. A \pill[blu]{promote} certificate ends the round in the \pill[lav]{Challenger's} favor: the \pill[lav]{Challenger} becomes the new \pill[lav]{Champion}, and the task opens the next round, or moves on to confirmation if its target is already met (\S\ref{sec:manager_evidence}). A \pill[ros]{reject} certificate fires when the result is negative or inconclusive. The \pill[lav]{Champion} stays in place, and the task opens a new round if budget remains, or halts. Not every certificate is a real choice. A guardrail breach or an unresolved integrity failure leaves only one legal response. The deterministic services then mark that response as forced, and the Manager's job narrows to confirming it or escalating. The facts a real decision would unlock stay withheld, since there is nothing left for them to decide.

\paragraph{Advancing to a new round.} A new round opens only after the previous round's effect judgment completes and the task still needs to search: the target is not yet met, and budget remains in $\mathcal{T}$. Opening it follows a fixed order. The Manager first writes the analysis specs that tell the deterministic services what to compute about the current search space. It then requests a plan from the Search Planner (\S\ref{sec:planner}); the Planner proposes candidates, and the Manager selects among them, never inventing candidates of its own. Each selected candidate becomes a \pill[lav]{Challenger}, a tree carrying one atomic action relative to the current \pill[lav]{Champion}. The Manager then launches the \pill[lav]{Challengers}, and the platform compiles and deploys them for the next observation window. Every round consumes traffic and round budget from $\mathcal{T}$, and the Manager tracks the remainder before opening the next. While a round runs, the Manager schedules the next look, marks a look diagnostic-only, extends the observation window, notifies stakeholders, and escalates on demand. It does not deploy a candidate or expand traffic without approval, alter a statistical conclusion, act on an action the deterministic services did not generate, or \pill[blu]{promote} a \pill[lav]{Champion} on search-period data alone.

\paragraph{Failure and stopping semantics.} The Manager does not always return a valid certificate. If its output is invalid or late, the system retries a bounded number of times and then falls back to a safe default, typically holding the round where it is or escalating to a human. Anything hard to reverse (deploying a candidate, expanding the traffic or round budget, or editing $\mathcal{T}$) always needs human approval; the Manager is never left to guess across that line. When a \pill[ros]{reject} leaves no budget in $\mathcal{T}$, the search stops. This means only that the task's rounds and traffic have run out---not that no further improvement is possible.

\subsection{Deterministic Services and Statistical System}
\label{sec:manager_services}

The lifecycle described in \S\ref{sec:manager_lifecycle} relies on the Experiment Manager to make judgment calls---which certificate to issue, when to escalate, how to interpret an anomaly. But every such call is bounded by a layer of deterministic services that the Manager cannot override (see the "Deterministic Services \& Statistical System" strip in Figure~\ref{fig:pilot_framework}). Six services form this layer, each with a narrow, auditable responsibility. Table~\ref{tab:det_services} summarizes each service, the artifact it produces, and the invariant it enforces. \textbf{The key separation: LLM roles hold judgment authority (choose among admissible options); deterministic services hold enforcement authority (generate the option space, validate every choice, commit state).}

\begin{table}[h]
\centering
\caption{Deterministic services bounding LLM judgment.}
\label{tab:det_services}
\small
\begin{tabularx}{\textwidth}{l l X}
\toprule
\textbf{Service} & \textbf{Artifact} & \textbf{Role} \\
\midrule
\rowcolor{gray!6}
Manager Guard & \texttt{Control Envelope} & Constructs admissible command set from state machine and frozen policy; validates the Manager's pick before commit \\
\addlinespace
\rowcolor{gray!6}
State Committer & \texttt{State Transition} & Sole state writer; atomically applies Guard-approved commands \\
\addlinespace
\rowcolor{teal!10}
Statistics Engine & \texttt{Decision Certificate} & Produces immutable per-look certificates; Manager may act on them but cannot alter conclusions \\
\addlinespace
\rowcolor{teal!10}
Contract Builder & \texttt{Learning Contract} & Freezes outcome criteria and execution plan before data collection begins \\
\addlinespace
\rowcolor{orange!10}
Candidate Validator & \texttt{Validation Report} & Per-action schema, envelope, and evidence checks on every \texttt{Planner Proposal} \\
\addlinespace
\rowcolor{orange!10}
Action Enumerator & \texttt{Action Envelope} & Lists all legal atomic actions for the current \pill[lav]{Champion} tree; Planner cannot invent actions outside it \\
\bottomrule
\end{tabularx}
\end{table}

\subsection{Confirmation and Delivery}
\label{sec:manager_evidence}

\paragraph{Staged scale-up.} A \pill[blu]{promote} certificate that also meets the task's target ends the search. From this point the Experiment Manager stops proposing new rounds and instead manages a staged handover from search traffic to production traffic. Scale-up proceeds through a sequence of stages. Each stage sets a traffic share, larger than the one before it, and a duration window with a minimum and a maximum. A stage waits at least its minimum before it can advance, and is re-evaluated by its maximum. At that point a certificate again resolves to \pill[blu]{hold}, \pill[blu]{advance}, or \pill[ros]{rollback}, using the same observation signals as in \S\ref{sec:manager_lifecycle}.

\paragraph{Independent confirmation and postmortem.} Scale-up alone does not confirm the \pill[lav]{Champion}. The data collected while searching, $D_{\text{search}}$, shaped the choice of \pill[lav]{Champion}, so it cannot also serve as proof. Final confirmation evaluates the \pill[lav]{Champion} against a single pre-registered test on $D_{\text{confirm}}$, a freshly collected, held-out sample gathered only after the \pill[lav]{Champion} was fixed. The test runs once. A result that is not significant on $D_{\text{confirm}}$ cannot be rescued by looking again or by folding in $D_{\text{search}}$. Once confirmed, the \pill[lav]{Champion} is delivered to production, and the task enters a postmortem. The Manager assembles a retrospective of what worked, what needed a retry or an escalation, and what the playbook $\mathcal{P}$ should record differently next time. This retrospective becomes a methodology patch consumed by the Memory Curator (\S\ref{sec:curator}). The task is not considered complete until the postmortem is resolved.

\section{Search Planner}
\label{sec:planner}

This section details the Search Planner: how the search problem is formulated (\S\ref{sec:planner_tree}), how candidates are pre-screened, explored, and ranked before committing experiment budget (\S\ref{sec:planner_prescreening}), and how the top-ranked candidates are verified through whole-bucket A/B testing in a multi-round loop (\S\ref{sec:planner_verification}).

\subsection{Problem Formulation}
\label{sec:planner_tree}

Table~\ref{tab:planner_notation} collects the notation used throughout this section and subsequent sections.

\begin{table}[t]
\centering
\caption{Notation used in the Search Planner and related sections.}
\label{tab:planner_notation}
\small
\begin{tabularx}{\textwidth}{l l X}
\toprule
\textbf{Symbol} & \textbf{Name} & \textbf{Description} \\
\midrule
$T$ & PolicyTree & A complete, executable mapping from the user population to strategy bundles \\
$T_0$ & Fixed baseline (\pill[lav]{B0}) & Platform-default tree, invariant across all rounds; anchors $\Delta_{\text{total}}$ \\
$T^*$ & \pill[lav]{Champion} & Current best confirmed tree; search reference within a reference epoch \\
$T^*_a$ & \pill[lav]{Challenger} & \pill[lav]{Champion} plus exactly one atomic action $a$; the unit of search \\
\midrule
$\mathcal{A}(T^*)$ & ActionEnvelope & Finite set of legal atomic actions for the current \pill[lav]{Champion}, enumerated deterministically \\
$a$ & Atomic action & One of five kinds: \textbf{split}, \textbf{prune}, \textbf{hashExpand}, \textbf{collapse}, \textbf{updateStrategy} \\
$\ell$ & Leaf & Terminal node of a PolicyTree; every user hits exactly one leaf \\
$b(\ell)$ & Bundle binding & Strategy bundle assigned to leaf $\ell$ \\
\midrule
$\mathcal{S}$ & SearchUniverse & Registered features $\mathcal{F}$, registered bundles, tree constraints, and search budget \\
$\mathcal{F}$ & Feature set & Pre-treatment registered features available for splits \\
$\mathcal{Y}(T)$ & Primary online metric & The observed value of the primary metric (e.g., IPV) under tree $T$ in the A/B platform \\
$w(\ell)$ & Global leaf share & $|\text{users in } \ell| \;/\; |\text{total users}|$ \\
$\Delta_{\text{total}}(T)$ & Total treatment effect & $\mathbb{E}[\mathcal{Y}(T)] - \mathbb{E}[\mathcal{Y}(T_0)]$; measured against the fixed baseline \\
$\Delta_{\text{step}}(a)$ & Step treatment effect & $\mathbb{E}[\mathcal{Y}(T^*_a)] - \mathbb{E}[\mathcal{Y}(T^*)]$; measured against the current \pill[lav]{Champion} \\
$\rho(a)$ & Resolvability & Whether action $a$ can be conclusively evaluated within the remaining budget \\
\bottomrule
\end{tabularx}
\end{table}

\begin{figure*}[t]
    \centering
    \includegraphics[width=0.83\linewidth]{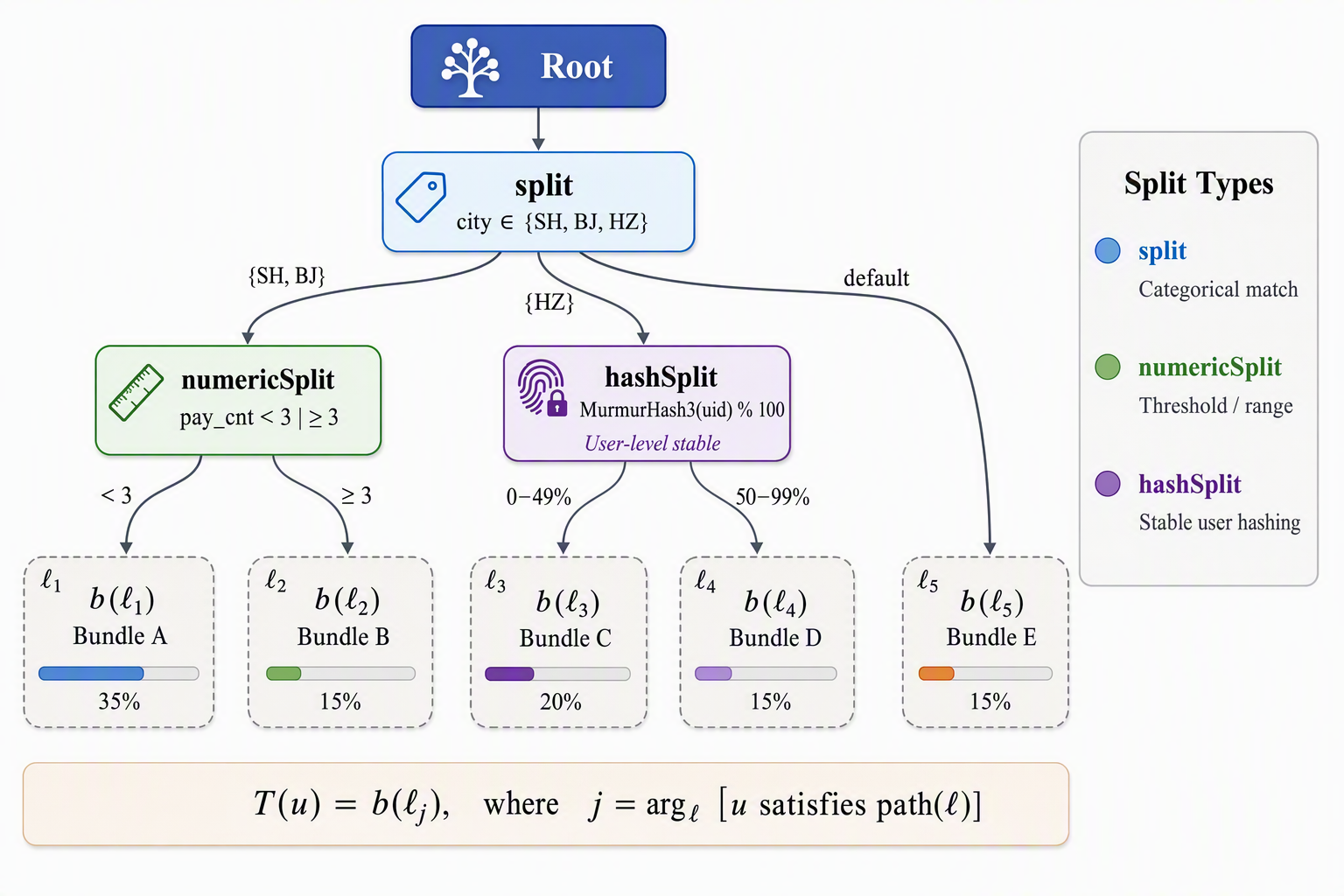}
    \captionsetup{justification=centering}
    \caption{Structure of a \texttt{PolicyTree}.}
    \label{fig:policytree_structure}
\end{figure*}

\paragraph{PolicyTree.}
A \texttt{PolicyTree} $T$ is a rooted tree that maps the entire user population to strategy bundles (Figure~\ref{fig:policytree_structure}). Every internal node routes users by a single pre-treatment registered feature $f \in \mathcal{F}$; every leaf $\ell$ binds exactly one strategy bundle $b(\ell)$. Formally, for a tree with leaves $\{\ell_1, \ldots, \ell_L\}$, the mapping can be written as
\begin{equation}
    T(u) \;=\; b(\ell_j), \quad \text{where } j = \arg\nolimits_{\ell} \bigl[u \in \text{path}(\ell)\bigr],
    \label{eq:tree_mapping}
\end{equation}
i.e., user $u$ is routed to the unique leaf whose path predicate it satisfies, and receives the strategy bundle bound to that leaf. Two structural invariants hold for every valid tree:
\begin{enumerate}
    \item \textbf{Completeness and mutual exclusion}: for any user $u$ in the population, there exists exactly one leaf $\ell$ such that $u$ satisfies the path predicate from root to $\ell$. The leaf predicates partition the population into disjoint, exhaustive segments---no user falls through, and no user matches two leaves.
    \item \textbf{Canonical identity}: semantically equivalent trees---those that route every user to the same bundle---produce the same canonical hash, regardless of node ordering or syntactic differences. This property ensures that the system can detect when two structurally different trees encode the same policy, preventing redundant experiments.
\end{enumerate}

\noindent Internal nodes come in three types that differ in how they partition users:
\begin{itemize}
    \item \textbf{split}: categorical feature match---routes users by an enumerated set of feature values (e.g., \texttt{city} $\in$ \{\texttt{Shanghai}, \texttt{Beijing}, \texttt{Hangzhou}\}), with a default child for unmatched values.
    \item \textbf{numericSplit}: numeric feature threshold or range---routes users by ordered comparisons ($<$, $\leq$, $>$, $\geq$, \texttt{between}), supporting both single-threshold and multi-interval boundaries.
    \item \textbf{hashSplit}: stable user-level hashing---computes a deterministic hash of a user identifier (via MurmurHash3) to map users to percentage buckets, enabling stable sub-population experimentation where the same user always lands in the same branch.
\end{itemize}

\paragraph{SearchUniverse.}
Before any search begins, the \texttt{SearchUniverse}~$\mathcal{S}$ defines the boundaries within which all trees must live. It contains: {\large\ding{182}}~the set of registered features $\mathcal{F}$, each with its data type, pre-treatment availability certification, missing-value semantics, and allowed values or boundaries; {\large\ding{183}}~the set of registered strategy bundles, each a versioned configuration that the online runtime can execute; {\large\ding{184}}~an initial tree $T^{(0)}$ that seeds the search; and {\large\ding{185}}~the hard tree constraints and search budget. The tree constraints are:
\begin{itemize}
    \item $d_{\max}$: maximum tree depth---limits the length of any root-to-leaf path, preventing overly specific population segments that lack statistical power.
    \item $\mathcal{L}_{\max}$: maximum leaf count---caps the total number of distinct strategy assignments, controlling deployment complexity.
    \item $w_{\min}$: minimum global leaf share---ensures every leaf receives enough traffic for the Statistics Engine to produce a conclusive \texttt{DecisionCertificate} within the budget.
    \item $\mathcal{F}_{\max}$: maximum distinct features---limits the number of different features used for splits across the entire tree, reducing the risk of overfitting to noise in high-dimensional feature spaces.
\end{itemize}
The search budget specifies the maximum number of rounds $\mathcal{R}_{\max}$, the total bucket-days available, the number of experiment buckets $\mathcal{K}$, and the maximum concurrent \pill[lav]{Challengers} per round. All these values are drawn from $\mathcal{S}$ and are not adjustable by the LLM; they are set by the user during task specification (\S\ref{sec:manager_intake}) and frozen before the search starts.

\paragraph{Search problem.}
The search begins from an initial tree $T^{(0)}$ specified in the \texttt{SearchUniverse}~$\mathcal{S}$ and seeks a sequence of trees $T^{(0)},\; T^{(1)},\; \ldots,\; T^{(\mathcal{R})}$, where each $T^{(r+1)}$ is obtained from $T^{(r)}$ by applying exactly one atomic action $a^{(r)} \in \mathcal{A}(T^{(r)})$. Let $\mathcal{Y}(T)$ denote the primary online metric (e.g., IPV) observed under the A/B platform when tree $T$ is deployed to its assigned traffic bucket. The optimization objective is:
\begin{equation}
    \begin{aligned}
        &\max_{a^{(0)},\ldots,a^{(\mathcal{R}-1)}} \;\; \Delta_{\text{total}}\bigl(T^{(\mathcal{R})}\bigr) \;=\; \mathbb{E}\bigl[\mathcal{Y}\bigl(T^{(\mathcal{R})}\bigr)\bigr] - \mathbb{E}\bigl[\mathcal{Y}(T_0)\bigr] \\
        &\text{s.t.} \quad
        \left\{
        \begin{aligned}
            \text{depth}(T^{(r)}) &\;\leq\; d_{\max}, \\
            \bigl|\text{leaves}(T^{(r)})\bigr| &\;\leq\; \mathcal{L}_{\max}, \\
            w(\ell) &\;\geq\; w_{\min} \;\; \forall\, \ell \in \text{leaves}(T^{(r)}), \\
            \bigl|\{f : f \text{ used in } T^{(r)}\}\bigr| &\;\leq\; \mathcal{F}_{\max},
        \end{aligned}
        \right.
        \\
    \end{aligned}
    \label{eq:search_objective}
\end{equation}
for all $r = 0, \ldots, \mathcal{R}$. These bounds---maximum tree depth $d_{\max}$, maximum leaf count $\mathcal{L}_{\max}$, minimum global leaf share $w_{\min}$, and maximum distinct features $\mathcal{F}_{\max}$---are drawn from $\mathcal{S}$ and are not adjustable by the LLM. They guarantee that every tree in the sequence is deployable online and that every leaf retains enough traffic for statistical resolution.

\paragraph{B0, Champion and Challenger.}
Three trees coexist during any round (Figure~\ref{fig:round_lifecycle}). The fixed baseline $T_0$ (\pill[lav]{B0}) runs the platform's default behavior throughout the experiment and never changes; it anchors $\Delta_{\text{total}}$. The \pill[lav]{Champion} $T^*$ is the current best confirmed tree and serves as the search reference within a reference epoch; every \pill[lav]{Challenger} is measured against it to obtain the step-level effect
\begin{equation}
    \Delta_{\text{step}}(a) \;=\; \mathbb{E}\bigl[\mathcal{Y}(T^*_a)\bigr] - \mathbb{E}\bigl[\mathcal{Y}(T^*)\bigr].
    \label{eq:delta_step}
\end{equation}
The \pill[lav]{Champion} is frozen within an epoch so that all \pill[lav]{Challengers} launched inside it share a common comparison. A \pill[blu]{promote} certificate (see \S\ref{sec:manager_lifecycle}) replaces the \pill[lav]{Champion} with the winning \pill[lav]{Challenger}; the old search data then serves only as historical evidence, not as the basis for promotion.

\paragraph{ActionEnvelope.}
Given the current \pill[lav]{Champion} $T^*$, the remaining depth budget, and the constraints in $\mathcal{S}$, the Action Enumerator (\S\ref{sec:manager_services}) deterministically constructs the \texttt{ActionEnvelope} $\mathcal{A}(T^*)$: the finite, auditable set of all atomic actions that can legally be applied to $T^*$. Each action $a \in \mathcal{A}(T^*)$ specifies a target leaf $\ell$, an action kind, and the resulting candidate tree $T^*_a$. The five action kinds are:
\begin{enumerate}
    \item \textbf{split}: partition a leaf on a categorical or numeric feature, creating child nodes with new bundle bindings.
    \item \textbf{prune}: merge an existing split, collapsing its children back into the parent leaf.
    \item \textbf{hashExpand}: add a \textbf{hashSplit} node for stable user-level sub-population experimentation.
    \item \textbf{collapse}: undo a structural expansion, merging an expanded subtree back.
    \item \textbf{updateStrategy}: change the bundle binding $b(\ell)$ on an existing leaf without altering tree structure.
\end{enumerate}
The Search Planner may only reference action IDs already present in $\mathcal{A}(T^*)$; it cannot invent actions outside the envelope. Actions not shortlisted by the Planner retain audit trails and remain eligible for the bounded exploration lane (\S\ref{sec:planner_preanalysis}).

\begin{figure*}[t]
    \centering
    \includegraphics[width=0.9\linewidth]{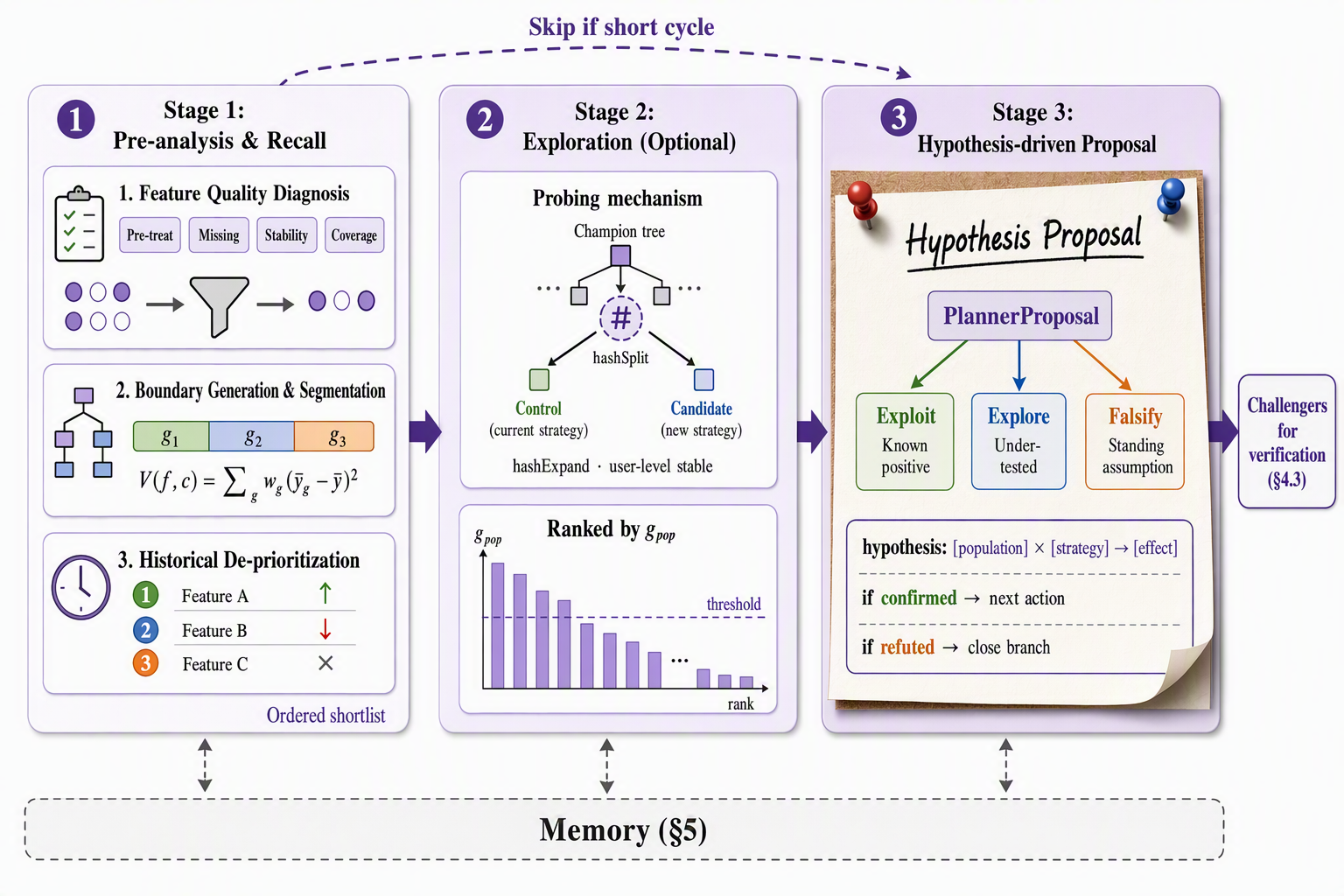}
    \caption{Search Planner pipeline: pre-analysis and candidate recall (Stage~1), optional exploration traffic via hashSplit probing (Stage~2), and hypothesis-driven proposal generation (Stage~3).}
    \label{fig:prescreening_pipeline}
\end{figure*}

\subsection{Pre-screening and Exploration}
\label{sec:planner_prescreening}

The search is organized into two temporal phases (Figure~\ref{fig:prescreening_pipeline}). \textbf{Round~0} completes pre-analysis and exploration traffic: it diagnoses feature quality, segments the population, runs small-scale within-population comparisons, and produces a priority-ranked candidate list. \textbf{Rounds~1--$N$} then verify candidates through whole-bucket A/B testing in an iterative loop (\S\ref{sec:planner_verification}). Round~0 results determine only the order in which candidates are verified; whether a candidate is ultimately adopted depends entirely on its whole-bucket A/B outcome.

\subsubsection{Pre-analysis and Candidate Recall}
\label{sec:planner_preanalysis}

Before any exploration traffic is allocated, the Search Planner runs a structured screening pipeline over the registered features in $\mathcal{F}$. The pipeline proceeds in three stages: quality diagnosis, boundary generation, and historical de-prioritization.

\paragraph{Feature-level quality diagnosis.}
For each registered feature $f \in \mathcal{F}$, the Planner checks four properties: {\large\ding{182}}~pre-treatment availability---whether $f$ can be observed before exposure, so that conditioning on it does not introduce post-treatment bias; {\large\ding{183}}~missing rate---features with excessive missing values produce unstable splits and reduce effective sample size; {\large\ding{184}}~temporal stability---whether the feature's distribution shifts significantly across days, which would confound repeated-measurement designs; and {\large\ding{185}}~coverage---whether the feature's non-missing population is large enough for each resulting subgroup to meet the minimum leaf share $w_{\min}$. Features that fail any hard threshold on these checks are excluded from the candidate set; those that pass enter the next stage with a quality score that influences their priority.

\paragraph{Candidate boundary generation and population segmentation.}
For numeric features that pass quality screening, the Planner generates candidate split boundaries from a deterministic set: distribution quantiles, known business-rule thresholds (e.g., membership tiers, purchase-count brackets), and historical points where strategy effects showed a change. For categorical features, the candidate set is the registered value enumeration. Each candidate boundary partitions the population into subgroups, and the Planner measures the exposure-weighted posterior variance---a proxy for how differently subgroups behave under the current policy:
\begin{equation}
    \mathcal{V}(f, c) \;=\; \sum_{g} w_g \;\bigl(\bar{y}_g - \bar{y}\bigr)^2,
    \label{eq:population_variance}
\end{equation}
where $g$ indexes the subgroups created by splitting feature $f$ at boundary $c$, $w_g$ is the exposure share of subgroup $g$, $\bar{y}_g$ is the subgroup posterior mean of the primary metric, and $\bar{y}$ is the overall posterior mean. A high $\mathcal{V}(f,c)$ indicates that subgroups under this split respond differently, making it a promising axis for personalization. Strongly correlated or near-duplicate features are identified and collapsed so that the downstream exploration budget is not spent on redundant comparisons.

\paragraph{Historical de-prioritization.}
The Planner cross-references the surviving feature--boundary pairs against historical evidence from Memory (\S\ref{sec:curator}). Feature--strategy combinations that failed to produce whole-bucket gain in prior experiments are deprioritized, and combinations whose mechanism has been explicitly falsified are dropped entirely. Conversely, combinations with positive but inconclusive prior signal are promoted, since they represent the lowest-cost opportunity for a decisive test. The output of this stage is an ordered shortlist of feature--boundary--strategy triples, ranked by a composite of population variance, quality score, and historical prior, ready for exploration traffic allocation.

\subsubsection{Exploration Traffic and Candidate Ranking}
\label{sec:planner_exploration}

This stage is not always executed. The Planner activates it only when two conditions hold simultaneously: {\large\ding{182}}~the remaining experiment cycle in $\mathcal{T}$ is long enough to absorb an extra probing round without squeezing the verification budget in Rounds~1--$N$, and {\large\ding{183}}~the number of surviving feature--boundary pairs from \S\ref{sec:planner_preanalysis} exceeds a threshold at which offline pre-analysis alone cannot reliably distinguish the most promising candidates. When the experiment cycle is short or the candidate set is already small, the Planner skips directly to three-lane candidate generation (\S\ref{sec:planner_three_lanes}), using the pre-analysis ranking as the sole priority signal.

\paragraph{Probing via expansion actions.}
When exploration is activated, the Planner deploys lightweight probing experiments by inserting \texttt{hashSplit} expansion nodes into the current \pill[lav]{Champion} tree. The \texttt{hashExpand} action from the \texttt{ActionEnvelope} serves this purpose: it inserts a \texttt{hashSplit} node that deterministically maps each user to a fixed branch via stable hashing. The same user always sees the same strategy across visits, enabling measurement of both immediate and cumulative effects (e.g., shifts in purchase frequency over a week). Because the assignment is user-level, each user contributes one independent observation per day, so sample-size accumulation tracks the number of exposed users rather than page views.

\paragraph{Candidate ranking.}
After the probing window closes, the Planner ranks candidates by their per-population gain contribution:
\begin{equation}
    g_{\text{pop}}(a, g) \;=\; w_g \;\cdot\; \bigl(\bar{y}^{\text{cand}}_g - \bar{y}^{\text{ctrl}}_g\bigr),
    \label{eq:pop_gain_contribution}
\end{equation}
where $\bar{y}^{\text{cand}}_g$ and $\bar{y}^{\text{ctrl}}_g$ are the per-user posterior metrics under the candidate and control strategies in subgroup $g$, and $w_g$ is the exposure share of subgroup $g$. Only results that meet two thresholds advance to the candidate list: {\large\ding{182}}~the accumulated sample size reaches the minimum required for a conclusive evaluation, and {\large\ding{183}}~the estimated effect is stable across at least the required number of consecutive observation days. Failing either gate disqualifies the result from influencing candidate priority, though it may still inform the Planner's understanding through the segment-diagnostic evidence role (\S\ref{sec:planner_verification}).

\paragraph{Global-share weighting.}
The per-population gains are aggregated into a single global contribution score for action $a$ by summing across all affected subgroups:
\begin{equation}
    g(a) \;=\; \sum_{g} g_{\text{pop}}(a, g) \;.
    \label{eq:global_contribution}
\end{equation}
This weighting ensures that a subgroup with extreme local lift but negligible population share does not dominate the ranking. The final candidate priority is determined by $g(a)$, so actions that improve large subgroups rank above those that produce large but narrow effects.

\begin{table}[t]
\centering
\caption{Bottom-up trajectory summarization vs.\ top-down hypothesis testing.}
\label{tab:bottomup_vs_topdown}
\small
\begin{tabularx}{\textwidth}{>{\columncolor{gray!8}}l >{\columncolor{green!8}}X >{\columncolor{orange!8}}X}
\toprule
& \textbf{Bottom-up (trajectory $\to$ summary)} & \textbf{Top-down (hypothesis $\to$ evidence)} \\
\midrule
Knowledge source & LLM post-hoc narrative over full trajectory & Pre-registered falsifiable hypothesis \\
\addlinespace
Direction control & Determined by LLM attention; uncontrollable & Determined by Planner proposal; auditable \\
\addlinespace
Confidence & No built-in measure; single-run and multi-run indistinguishable & Grounded in statistical certificate; confidence = design power \\
\addlinespace
Negative results & Typically discarded or vaguely noted & Explicitly close a hypothesis branch; prevent re-exploration \\
\addlinespace
Hallucination risk & High---narrative coherence $\neq$ empirical validity & Low---every claim has a provenance chain \\
\bottomrule
\end{tabularx}
\end{table}

\subsubsection{Hypothesis-driven Candidate Proposal}
\label{sec:planner_three_lanes}

\paragraph{Motivation.}
Most existing LLM-driven optimization frameworks accumulate experience through a bottom-up pattern: the system collects a trajectory of experimental actions and outcomes, then asks the LLM to summarize the trajectory into reusable heuristics or rules~\cite{agenticrectune2026,agentx2026,evorec2026}. While straightforward, this paradigm has three structural weaknesses:

\begin{enumerate}
    \item \textbf{High stochasticity.} The summarization step relies entirely on the LLM's post-hoc reasoning over noisy, high-dimensional trajectories. Different summarization runs on the same trajectory can yield contradictory conclusions.
    \item \textbf{Uncontrollable accumulation direction.} Because the LLM decides \textbf{after the fact} which aspects of the trajectory matter, the system has no control over what knowledge is retained; experience grows along whichever axis the LLM happens to attend to.
    \item \textbf{No built-in confidence.} A heuristic distilled from a single lucky run is indistinguishable from one confirmed across dozens of experiments, creating a fertile ground for hallucinated insights that propagate through subsequent decisions.
\end{enumerate}

\noindent PILOT replaces this bottom-up pattern with a \textbf{top-down, hypothesis-first} mechanism. Before any experiment budget is spent, the Search Planner is required to commit a falsifiable hypothesis---specifying the target population, the strategy change, the expected causal mechanism, and the criteria under which the hypothesis would be considered refuted. Evidence is then collected \textbf{specifically to test that hypothesis} through pre-registered A/B comparison. A hypothesis that survives verification is persisted as a high-confidence experience record with its supporting evidence; one that is refuted is persisted as a negative result that closes a branch of the search space. In both cases, the experience carries an auditable provenance chain (hypothesis $\to$ contract $\to$ data $\to$ certificate), and its confidence is determined by the statistical design rather than by the LLM's narrative coherence. Table~\ref{tab:bottomup_vs_topdown} summarizes the comparison.

\definecolor{hyp_bg}{RGB}{248, 248, 255}
\definecolor{hyp_frame}{RGB}{75, 105, 160}
\definecolor{hyp_field}{RGB}{245, 248, 255}
\definecolor{hyp_outcome_pos}{RGB}{240, 255, 240}
\definecolor{hyp_outcome_neg}{HTML}{CE1F49}

\begin{tcolorbox}[
enhanced,
colback=hyp_bg,
colframe=hyp_frame,
boxrule=1.5pt,
arc=3mm,
width=\textwidth,
title={\textbf{Case: A \texttt{PlannerProposal} in Round~2}},
coltitle=white,
fonttitle=\bfseries\small,
attach boxed title to top left={yshift=-3mm, xshift=6mm},
boxed title style={colback=hyp_frame, arc=2mm},
breakable
]
\small

\textbf{Context.}
Round~1 applied \textbf{split} on \texttt{purchase\_count} at boundary $\geq 3$, producing two leaves. The whole-bucket A/B result showed a positive $\Delta_{\text{step}}$ that met the promotion threshold; the \pill[lav]{Challenger} was promoted to \pill[lav]{Champion$_1$}. Segment-diagnostic evidence revealed that within the high-purchase leaf ($\texttt{purchase\_count} \geq 3$), users with $\texttt{visit\_frequency} \geq 5$/week responded $2.3\times$ more strongly than others, but this observation carries only diagnostic weight and cannot trigger promotion on its own.

\vspace{0.3cm}
\colorbox{hyp_field}{
\begin{minipage}{0.95\linewidth}
\textbf{ActionId}: \texttt{split\_leaf\_03\_on\_visit\_frequency\_ge5}\\[2pt]
\textbf{Ordinal Priority}: 1 {\small (exploit)}\\[2pt]
\textbf{Hypothesis}: ``Splitting the high-purchase leaf (\texttt{purchase\_count} $\geq 3$) by \texttt{visit\_frequency} at threshold 5 and assigning the enhanced ranking formula \texttt{S2} to the high-frequency subgroup will improve whole-bucket IPV. \textit{Reject if}: the whole-bucket $\Delta_{\text{step}}$ is non-positive after the full observation window, or if the GMV guardrail declines by more than 0.5\%.''\\[2pt]
\textbf{Evidence Refs}: \texttt{[round\_1\_certificate, segment\_diag\_visit\_freq\_heterogeneity]}\\[2pt]
\textbf{Expected Mechanism}: High-frequency visitors in the high-purchase segment have more exposure opportunities per unit time; a ranking formula optimized for repeat engagement (\texttt{S2}) should convert these exposures more efficiently than the generic formula (\texttt{S0}).
\end{minipage}
}

\vspace{0.3cm}
\textbf{Outcome for Next Decision}:

\vspace{0.15cm}
\colorbox{hyp_outcome_pos}{
\begin{minipage}{0.95\linewidth}
\textbf{If confirmed} $\to$ The split on \texttt{visit\_frequency} is validated. Next round: explore whether a further \textbf{numericSplit} on \texttt{avg\_session\_duration} within the high-frequency subgroup can capture an additional effect. Persist: ``\texttt{visit\_frequency $\geq$ 5} $\times$ \texttt{S2} is positive for high-purchase users.''
\end{minipage}
}

\vspace{0.15cm}
\colorbox{hyp_outcome_neg!12}{
\begin{minipage}{0.95\linewidth}
\textbf{If refuted} $\to$ Close the \texttt{visit\_frequency} branch for this population. The segment-diagnostic heterogeneity did not survive whole-bucket validation. Persist: ``\texttt{visit\_frequency} split does not produce whole-bucket gain in the high-purchase segment; do not re-prioritize.'' Next round: investigate the second-ranked hypothesis (alternative strategy \texttt{S3} on the same leaf without further splitting).
\end{minipage}
}
\end{tcolorbox}

\paragraph{Hypothesis formulation.}
When the Experiment Manager issues a \texttt{RequestPlan}, the Planner reads the current \pill[lav]{Champion}, the \texttt{ActionEnvelope} $\mathcal{A}(T^*)$, all accumulated evidence (pre-analysis results, exploration signals if available, and outcomes of prior verification rounds), and relevant entries from Memory (\S\ref{sec:curator}). From this input, the Planner performs four reasoning steps: {\large\ding{182}}~review the previous round's hypothesis outcome---supported, refuted, or still undetermined; {\large\ding{183}}~identify the uncertainty that most limits the next decision; {\large\ding{184}}~select actions from $\mathcal{A}(T^*)$ that either \textbf{exploit} a known positive signal, \textbf{explore} an under-tested region, or \textbf{falsify} a standing assumption; and {\large\ding{185}}~for each selected action, specify how different experimental outcomes would change subsequent search direction. The exploit--explore--falsify trichotomy ensures that the Planner's proposals are not uniformly greedy: falsification proposals are valued precisely because a negative result permanently prunes a hypothesis branch, accelerating convergence even when no positive gain is found.

\paragraph{PlannerProposal.}
The Planner's sole formal output is a \texttt{PlannerProposal}. Each proposed action carries six fields:
\begin{itemize}
    \item \textbf{ActionId}: must exist in the current $\mathcal{A}(T^*)$.
    \item \textbf{Ordinal Priority}: the Planner's ranking within its own shortlist.
    \item \textbf{Hypothesis}: a falsifiable statement specifying the target population, the strategy change, the expected direction, and the conditions under which the hypothesis would be rejected.
    \item \textbf{Evidence Refs}: pointers to evidence entries that support the hypothesis.
    \item \textbf{Expected Mechanism}: a causal sketch of how the action produces its effect.
    \item \textbf{Outcome for Next Decision}: a conditional plan---if the hypothesis is confirmed, which follow-up action to pursue; if refuted, which hypothesis branch to close and what alternative to investigate.
\end{itemize}

The proposal must not contain treatment-effect estimates, standard errors, significance levels, posterior probabilities, traffic-allocation recommendations, promotion decisions, observation-window adjustments, or early-stop conclusions---these belong to the Statistics Engine and the Experiment Manager, not to the Planner. This separation ensures that the LLM's contribution is limited to \textbf{what to test and why}, while \textbf{whether the test succeeded} is determined entirely by the deterministic statistical machinery.

\subsection{Candidate Verification}
\label{sec:planner_verification}
Once pre-screening produces a ranked candidate list, the search enters an iterative verification loop (Rounds~1--$N$). Each round converts one or more top-ranked candidates into \pill[lav]{Challenger} trees---each differing from the current \pill[lav]{Champion} by exactly one atomic action---and evaluates them through whole-bucket A/B testing. The key design principle is that \textbf{no candidate is adopted based on pre-screening evidence alone}; adoption requires a statistically conclusive whole-bucket comparison against the \pill[lav]{Champion}.

\paragraph{Pre-registered decision criteria.}
Before a \pill[lav]{Challenger} is deployed, the system freezes the decision criteria for that round: the hypothesis to be tested, the primary metric, the minimum effect size worth detecting, guardrail thresholds, and the rules that map each possible statistical outcome (positive, negative, inconclusive, or extend observation) to a concrete next action. Freezing these criteria before data collection prevents the common pitfall of post-hoc rationalization---adjusting what counts as ``success'' after seeing the results. The Planner specifies \textbf{what} to test; the deterministic statistical machinery decides \textbf{whether} it passed.

\paragraph{Verification loop.}
Each round proceeds as follows. The \pill[lav]{Challengers} are compiled into executable policy trees and deployed to the A/B platform alongside the \pill[lav]{Champion}. During the observation window, the Experiment Manager collects metrics and monitors data health (\S\ref{sec:manager_lifecycle}). At the close of observation, the statistical system produces a verdict: \pill[blu]{promote} (the \pill[lav]{Challenger} becomes the new \pill[lav]{Champion}), \pill[ros]{reject} (the \pill[lav]{Champion} is retained), or \pill[blu]{continue} (the observation window is extended). When multiple \pill[lav]{Challenger}s run in parallel within the same round, statistical thresholds are jointly calibrated across the multi-arm comparison to control the overall error rate. After each verdict, the outcome is persisted as a strategy experience record---including the hypothesis, the target population, the metric result, and whether the whole-bucket threshold was met. Confirmed hypotheses become high-confidence positive experience; refuted hypotheses become negative experience that prevents future re-exploration of the same candidate. Both feed into the Memory Curator (\S\ref{sec:curator}), ensuring that subsequent rounds---and future experiments on the same search universe---benefit from accumulated evidence rather than starting from scratch.
\section{Memory Curator}
\label{sec:curator}

The Memory Curator transforms completed experiments and subsequent user feedback into reusable, cross-experiment knowledge. It maintains two complementary memory stores. The first, \textbf{Strategy Evidence}, organizes evidence around strategy--population bundles. Each entry associates a strategy with its target population, the resulting online metrics, and the observed positive or negative outcomes. The second store, \textbf{Methodology Experience}, captures operational knowledge about how experiments should be conducted. It records concise summaries of actions such as waiting, retrying, preserving traffic or sample volume, extending observation, ramping up traffic, and rolling back an intervention, together with their execution context, outcomes, and relevant user feedback. When new feedback is received, the Memory Curator reconciles operational records across experiments, analyzes the likely causes of recurring successes or failures, and uses the resulting evidence to refine the corresponding operational skills.

A key design boundary separates \textit{task-level statistical conclusions} from \textit{cross-task memory confidence}. A terminal certificate issued by the Statistics Engine---\pill[blu]{promote}, \pill[ros]{reject}, or \texttt{inconclusive}---determines the next action for the current task: replacing the \pill[lav]{Champion}, retaining it, or extending the experimental round. However, even a decisive certificate can create only a \pill[ros]{draft} memory entry. No single experiment is sufficient to establish reusable cross-task knowledge. Advancing an entry from \pill[ros]{draft} to \pill[blu]{supported} and ultimately to \pill[blu]{approved} requires consistent evidence from independent tasks, as assessed through the reconciliation process (\S\ref{sec:curator_reconciliation}).

Memory remains advisory at every confidence level. It may influence strategy recall, candidate prioritization, experimental planning, and the selection or refinement of operational methods, but it never substitutes for a live \texttt{DecisionCertificate}. In particular, memory cannot directly justify a promotion, generate statistical evidence, or create or expand control permissions.

Table~\ref{tab:two_stores} summarizes the two memory stores. Section~\ref{sec:curator_forkmerge} introduces their shared fork--merge architecture, while \S\ref{sec:curator_reconciliation} defines the reconciliation procedure and confidence lifecycle used by both stores. Sections~\ref{sec:curator_bundle} and \ref{sec:curator_methodology} describe Strategy Evidence and Methodology Experience, respectively. Finally, \S\ref{sec:curator_governance} specifies the write-governance rules and the self-improvement flywheel enabled by this architecture.

\begin{table}[t]
\centering
\caption{Comparison of the two memory stores maintained by the Memory Curator.}
\label{tab:two_stores}
\small
\begin{tabularx}{\textwidth}{l X X}
\toprule
& \textbf{Strategy Evidence} & \textbf{Methodology Experience} \\
\midrule
\rowcolor{gray!6}
Scope & Domain knowledge: which bundle outperforms which comparator, on which population & Meta-knowledge: how to run experiments well under specific operational scenarios \\
\addlinespace
Retrieval key & \texttt{(from\_bundle\_id, to\_bundle\_id, population\_scope)} & \texttt{(operation\_type, scenario\_predicate)} \\
\addlinespace
\rowcolor{gray!6}
Data model & \texttt{EvidenceItem} (immutable per-round observations) aggregated into \texttt{ScopedClaim} (cross-round conclusions) & \texttt{PostmortemRecord} (immutable per-task retrospectives) distilled into \texttt{MethodologyEntry} (evolving operational recommendations) \\
\addlinespace
Written when & A round closes with a \pill[blu]{promote}, \pill[ros]{reject}, or \texttt{inconclusive} certificate & The task postmortem produces a methodology patch \\
\addlinespace
\rowcolor{gray!6}
Consumed by & Search Planner: candidate recall and historical de-prioritization (\S\ref{sec:planner_preanalysis}) & Experiment Manager: playbook assembly and runtime decisions (\S\ref{sec:manager_intake}) \\
\addlinespace
Confidence driver & Consistent \texttt{ScopedClaim} conclusions across independent tasks & Consistent operational outcomes across independent tasks \\
\bottomrule
\end{tabularx}
\end{table}

\subsection{Memory Architecture and Fork-Merge}
\label{sec:curator_forkmerge}

The two stores serve complementary roles within the experiment loop. Strategy Evidence (\S\ref{sec:curator_bundle}) records \textit{what experiments discovered}: every entry contrasts one strategy bundle against another on a defined population, tagged with whether the observation supports a causal claim or only a diagnostic one. Methodology Experience (\S\ref{sec:curator_methodology}) records \textit{how to run experiments}: observation windows, retry policies, scale-up ladders, rollback procedures, and monitoring cadences, each tracked with source provenance and a governed confidence level.

\begin{figure*}[t]
\centering
\includegraphics[width=0.8\linewidth]{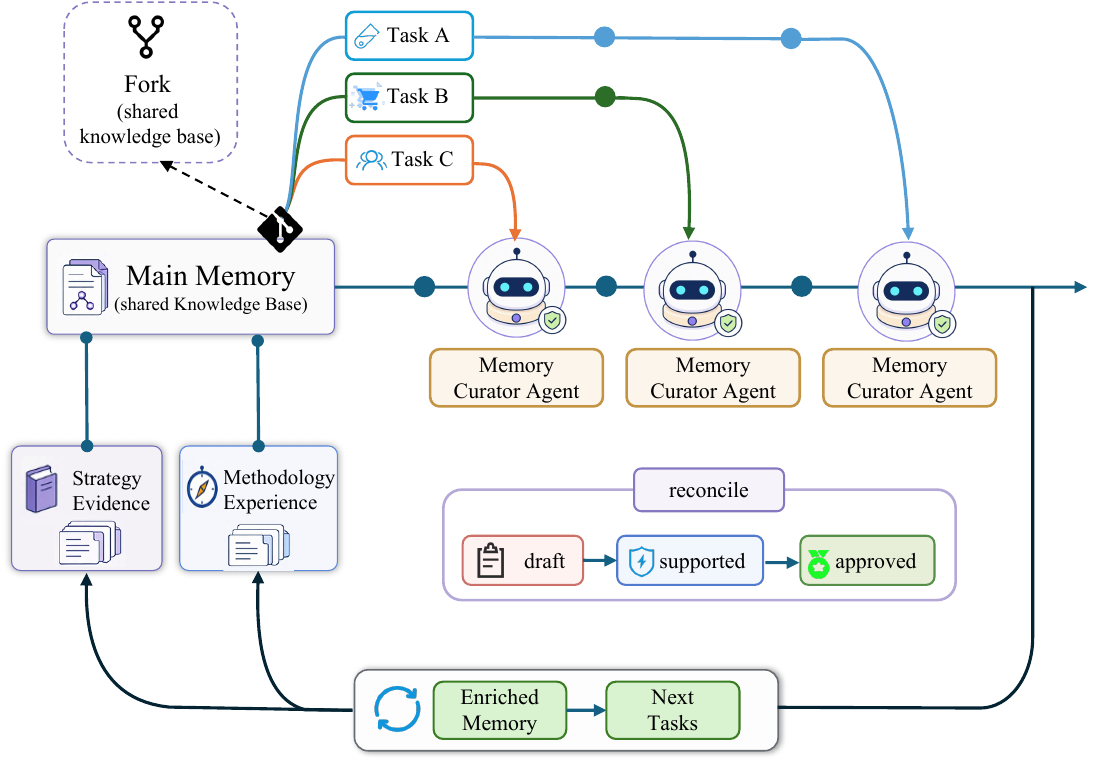}
\caption{Memory fork--merge lifecycle in \pilot{}. The Curator forks an isolated branch from the shared main memory for each task, covering both Strategy Evidence and Methodology Experience. Task branches remain invisible to other concurrent tasks during execution. After terminal settlement, the Memory Curator Agent reconciles the branch with the main store, advances claim confidence through the governed lifecycle, and merges approved updates back into main memory for use by subsequent tasks.}
\label{fig:memory_fork_merge}
\end{figure*}

When a new task begins, the Curator forks an isolated branch from the main knowledge base. The branch spans both stores: all \texttt{EvidenceItem}s, \texttt{ScopedClaim}s, and \texttt{MethodologyPatch}es produced during the task are deposited into this branch and remain invisible to other concurrent tasks. If two tasks run in parallel, each writes to its own branch; their results converge only at merge time, never during execution.

Figure~\ref{fig:memory_fork_merge} illustrates this lifecycle. On the left, the main memory---containing both Strategy Evidence and Methodology Experience---serves as the shared knowledge base. Each new task (A, B, C, \ldots) forks a local branch from this base; the branches run concurrently and in isolation, with no cross-visibility during execution. When a task's evidence reaches terminal settlement, the Memory Curator Agent takes over: it verifies that the evidence is settled, reconciles the branch entries against the main store using the retrieval-key matching and three-case logic detailed in \S\ref{sec:curator_reconciliation}, and merges the approved updates back into the main memory. The confidence lifecycle (\pill[ros]{draft} $\to$ \pill[blu]{supported} $\to$ \pill[blu]{approved}) advances only through this cross-task reconciliation, and the updated main memory is reused by the next task, closing the compounding loop.

The branch enters the merge process only after the task completes and its evidence reaches terminal settlement (\S\ref{sec:manager_evidence}). At that point, the Curator applies the reconciliation logic described in \S\ref{sec:curator_reconciliation} to integrate the branch entries into the main stores. Should the merge itself fail---due to a reconciliation conflict the Curator cannot resolve, an LLM timeout, or any other curation error---the already-settled evidence, the Manager's current decision, and the next Planning request all remain valid. Failure degrades knowledge accumulation for future tasks; it does not compromise the correctness of the task that produced the evidence.

\subsection{Reconciliation and Confidence Lifecycle}
\label{sec:curator_reconciliation}

The Memory Curator assigns each memory entry one of three confidence states:
\pill[ros]{draft}, \pill[blu]{supported}, or \pill[blu]{approved}.
These states quantify the degree of cross-task validation received by a claim, rather than the statistical strength of the experiment that initially produced it.

\begin{itemize}
    \item \pill[ros]{draft} denotes a task-local observation that has been admitted to the shared memory but has not yet been independently replicated. Every newly created entry begins in this state, regardless of the sample size or statistical significance of its originating experiment.

    \item \pill[blu]{supported} denotes a claim that has received at least one consistent confirmation from an independent task under the same retrieval scope. Such an entry represents a recurring observation, but the available evidence remains insufficient for unrestricted reuse.

    \item \pill[blu]{approved} denotes a claim whose supporting evidence has exceeded a governed cross-task confirmation threshold without unresolved material contradictions. Approved entries may be used autonomously for advisory functions such as strategy recall, candidate prioritization, playbook assembly, and operational-method selection. 
\end{itemize}

The nominal confidence progression is
\[
\pill[ros]{draft}
\;\longrightarrow\;
\pill[blu]{supported}
\;\longrightarrow\;
\pill[blu]{approved}.
\]
Confidence advances only when a claim is consistently confirmed by independent tasks. Repeated observations or evaluations derived from the same task count as a single source of evidence, and duplicate records provide no additional support. When contradictory evidence arises, the Curator re-evaluates the claim and may suspend its advancement or mark it as superseded if the opposing evidence becomes sufficiently strong. Every confidence transition and supersession is recorded together with its evidence and rationale; existing claims are never silently overwritten.

\paragraph{Reconciliation procedure.}
Each task writes new observations to an isolated memory branch. After the task reaches terminal settlement, the Curator reconciles each branch entry against the corresponding main store. It retrieves potentially related entries using a store-specific key:
\texttt{(from\_bundle\_id, to\_bundle\_id, population\_scope)}
for Strategy Evidence, and
\texttt{(operation\_type, scenari} \texttt{o\_predicate)}
for Methodology Experience.
For Strategy Evidence, the reconciled conclusion describes the online effect of a strategy transition for a specified population. For Methodology Experience, it describes the observed utility or failure mode of an operational method in a specified scenario, incorporating experimental outcomes and relevant user feedback.

For each incoming entry, reconciliation produces one of three outcomes:

\begin{enumerate}
    \item \textbf{Consistent evidence.}
    If the incoming conclusion agrees with an existing claim under the same retrieval key, the Curator appends the task as a supporting reference. The first qualifying independent confirmation advances the entry from \pill[ros]{draft} to \pill[blu]{supported}. Once the accumulated independent confirmations satisfy the governed approval criterion, the entry advances from \pill[blu]{supported} to \pill[blu]{approved}. Further consistent observations strengthen the evidence base of an already-approved entry without changing its state.

    \item \textbf{Contradictory evidence.}
    If the incoming conclusion conflicts with an existing claim under the same key, the Curator records it as a contradicting reference and re-evaluates the claim's confidence. An isolated contradiction does not automatically erase or reverse the existing entry. Instead, the Curator preserves both the supporting and contradicting evidence, prevents unsupported confidence advancement, and applies the governed conflict-resolution rule.

    If independent contradictory evidence becomes sufficiently strong to overturn the existing claim, the original entry is explicitly marked as superseded, together with the evidence and rationale for that decision. The competing claim is materialized as a new \pill[ros]{draft} entry and must independently progress through the same cross-task validation lifecycle. Thus, even strong counter-evidence cannot immediately create an approved replacement.

    \item \textbf{No matching claim.}
    If no existing entry shares the retrieval key, the incoming entry is inserted into the main store as \pill[ros]{draft}. It remains a provisional observation until subsequent independent tasks provide supporting or contradicting evidence.
\end{enumerate}

\paragraph{Cross-task confidence versus within-task significance.}
The confidence lifecycle is deliberately separated from task-level statistical inference. The Statistics Engine determines whether an observed effect is sufficiently supported \emph{within} the current experiment and issues a terminal
\texttt{DecisionCertificate}. The Memory Curator instead determines whether the resulting claim is reliable \emph{across} experiments. Consequently, a single task can produce only a \pill[ros]{draft} entry, even when its sample size is large and its statistical conclusion is decisive. This separation prevents a one-off result from being prematurely promoted into generally reusable knowledge.

A terminal \texttt{inconclusive} result is treated as a distinct semantic outcome rather than as negative evidence. It indicates that the task lacked sufficient information or statistical power to resolve the claim. The Curator may retain the associated execution record for diagnostic purposes, but it neither counts the result as a contradiction nor changes the confidence of an existing claim.

\paragraph{Scope compatibility.}
Confidence transitions require exact compatibility of the retrieval scope. In Strategy Evidence, this includes the strategy transition and population predicate; in Methodology Experience, it includes the operation type and scenario predicate. Entries with overlapping but non-identical scopes may be retrieved as contextual evidence, but they do not directly drive confidence transitions.

For example, evidence obtained for
\texttt{purchase\_count $\geq 3$ AND visit\_freq $\geq 5$}
may inform the interpretation of a claim for
\texttt{purchase\_count $\geq 3$},
but it cannot independently confirm or contradict that broader claim. This restriction prevents evidence from a narrow subpopulation or operational scenario from being incorrectly generalized to a broader scope. Where recurring scope-specific differences are observed, the Curator retains separate entries rather than forcing them into a single aggregate claim.

\subsection{Strategy Evidence}
\label{sec:curator_bundle}

Strategy Evidence records cross-task knowledge about the relative effectiveness of strategy bundles: which bundle performs better than which baseline, for which population, and with what confidence. After an experiment round terminates with a certificate (\S\ref{sec:manager_lifecycle}), the Curator
stores its result and reconciles it with evidence from prior tasks using the procedure in \S\ref{sec:curator_reconciliation}. The Search Planner uses this knowledge to recall promising candidates and de-prioritize historically ineffective ones (\S\ref{sec:planner_preanalysis}), while the Experiment
Manager uses it to identify relevant historical evidence for the current task. An illustrative Strategy Evidence entry is provided in
Appendix~\ref{app:memory_cases}.

\paragraph{Bundle identity and pairwise contrast.}
A strategy bundle comprises the strategies applied to users in an experiment and the configuration governing their execution. A versioned \texttt{StrategyBundleSpec} records the strategy list, execution order, eligibility rules, trigger rules, and shared-resource policies. The Curator derives a canonical \texttt{strategy\_bundle\_id} from this specification, enabling bundles with the same configuration to be identified consistently across tasks.

Each observation records a directed comparison between two bundles: \texttt{from\_bundle\_id} identifies the baseline bundle, and \texttt{to\_bundle\_id} identifies the evaluated bundle. It also records \texttt{population\_scope}, the user condition to which the comparison applies. An effect such as ``IPV increased by 3\%'' is therefore stored together with both its baseline and target population. This representation allows later tasks to retrieve evidence for the same comparison without conflating results obtained against different baselines or on different populations.

Only observations with the same bundle comparison and population scope can jointly advance a conclusion's confidence. Evidence from an overlapping but non-identical population may inform a follow-up experiment, but it does not count as an independent confirmation.

\paragraph{Immutable observations and evolving conclusions.}
Strategy Evidence separates individual experiment results from conclusions formed across tasks. An \texttt{EvidenceItem} is an immutable record of one experiment round, containing the bundle comparison, population scope, online metric outcomes, and statistical certificate. Once stored, it is never modified.

A \texttt{ScopedClaim} is a cross-task conclusion that aggregates \texttt{EvidenceItem}s with the same bundle comparison and population scope. It retains both supporting and contradicting observations, and its confidence follows the lifecycle defined in \S\ref{sec:curator_reconciliation}. If sufficient contradictory evidence overturns a claim, the Curator marks it as superseded and records the reason rather than overwriting it. Thus, conclusions can evolve while the underlying experimental records remain intact.

\paragraph{Causal and diagnostic evidence.}
The Curator distinguishes evidence by how it was produced. \textit{Causal evidence} comes from a pre-specified, randomized A/B comparison between complete strategy bundles. It can support a \texttt{ScopedClaim} and advance its cross-task confidence. \textit{Diagnostic evidence} comes from post-hoc analyses, such as within-group feature slices or expansion probes (\S\ref{sec:planner_exploration}). It may reveal population differences and help the Search Planner propose follow-up candidates, but it cannot advance a claim's confidence or be treated as a validated causal result.

When a \pill[lav]{Challenger} changes multiple strategies relative to the \pill[lav]{Champion}, the Curator records only the effect of the complete bundle. It does not attribute the observed effect to individual strategies because the experiment cannot identify their separate contributions. If such attribution is needed, the Search Planner may propose subsequent rounds that change one strategy at a time.

\paragraph{Use and boundaries.}
The Search Planner queries Strategy Evidence for prior results, unresolved comparisons, and contradictory observations. Prior results inform candidate recall and prioritization, unresolved comparisons motivate further testing, and recurring contradictions may indicate that the population scope should be refined.

Evidence from the current round becomes available as memory only after the round terminates and reconciliation completes; it cannot re-enter the same round as historical evidence. If a remembered conclusion conflicts with a live A/B result, the live \texttt{DecisionCertificate} always takes precedence. Strategy Evidence informs candidate selection and experiment planning, but it never replaces current statistical evidence or directly determines promotion.

\begin{figure*}[t]
\centering
\includegraphics[width=0.9\linewidth]{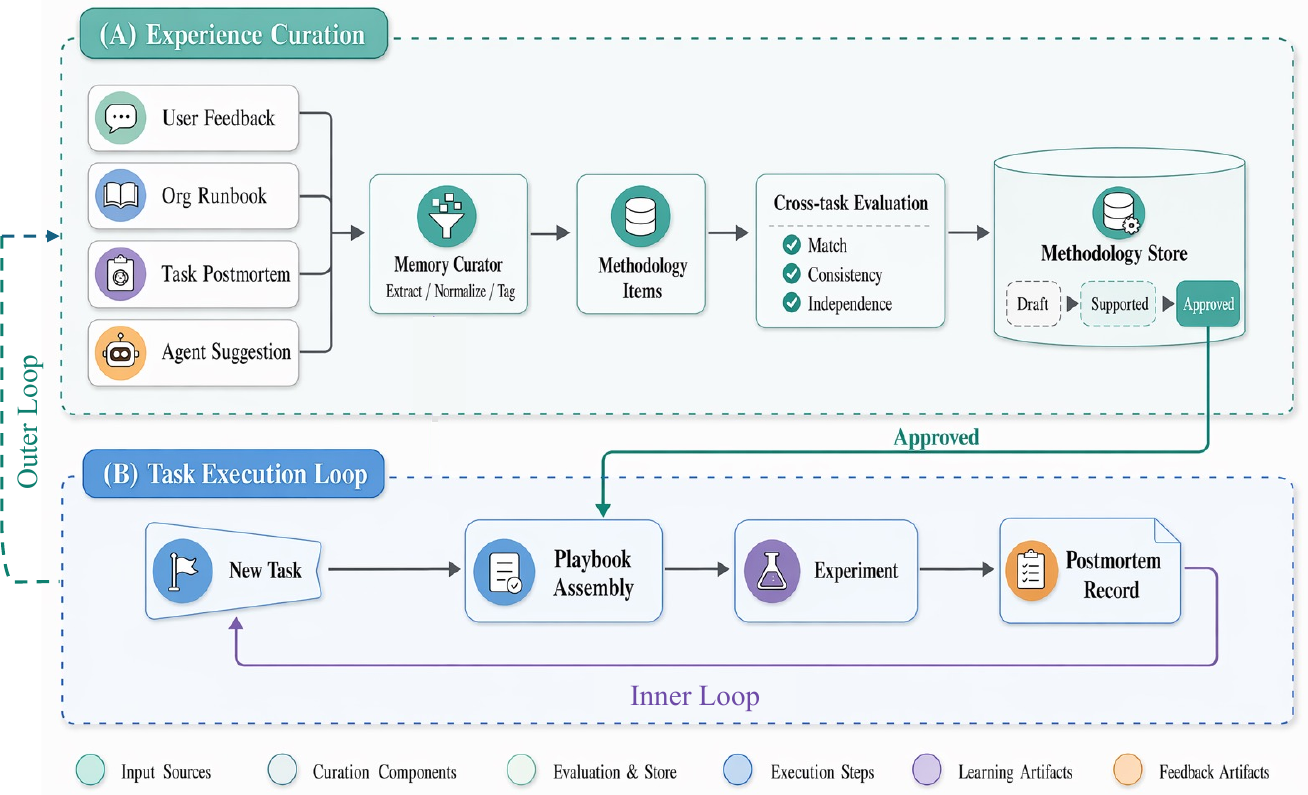}
\caption{Methodology Experience lifecycle in \pilot{}. The experience-curation path (A) extracts recommendations from user feedback, organizational runbooks, task postmortems, and agent suggestions, evaluates them across tasks, and stores them with confidence states from \pill[ros]{draft} to \pill[blu]{supported} and \pill[blu]{approved}. The task-execution loop (B) reuses approved methodology to assemble playbooks, run experiments, and generate postmortem records, which feed back into future curation.}
\label{fig:methodology_experience}
\end{figure*}

\subsection{Methodology Experience}
\label{sec:curator_methodology}

While Strategy Evidence records what works, Methodology Experience records how to run experiments effectively. It captures operational practices for waiting, retrying, guaranteeing traffic volume, observing a running experiment, scaling up traffic, and rolling back. After each task, the Experiment Manager produces a postmortem (\S\ref{sec:manager_evidence}) that summarizes the operations
performed, the problems encountered, and the procedures that should be retained or adjusted. The Memory Curator combines this postmortem with explicit user feedback to create \texttt{MethodologyEntry}s, which are operational recommendations that can be reused by future tasks. An illustrative Methodology Experience entry is provided in Appendix~\ref{app:memory_cases}.

Figure~\ref{fig:methodology_experience} shows this lifecycle. In the experience-curation path (A), the Curator extracts recommendations from experiment records, postmortems, and user feedback, and reconciles them with methodology from prior tasks. Each new recommendation begins as \pill[ros]{draft} and advances through \pill[blu]{supported} to \pill[blu]{approved} only when independent tasks provide consistent evidence, following \S\ref{sec:curator_reconciliation}. In the task-execution loop (B), the Experiment Manager retrieves relevant \pill[blu]{approved} entries to assemble a playbook for a new task. The resulting experiment produces another postmortem, which returns to the curation path. Each task therefore both uses and extends the operational knowledge base.

\paragraph{Operation categories and retrieval.}
Each \texttt{MethodologyEntry} is indexed by an \texttt{operation\_type}, which identifies the operational decision it addresses, and a \texttt{scenario\_predicate}, which specifies when the recommendation applies. The six operation types cover the main decisions made during experiment execution:

\begin{itemize}
    \item \textbf{wait}: determine how long to collect evidence before the next decision;
    \item \textbf{retry}: recover from an anomaly or an inconclusive round;
    \item \textbf{guarantee volume}: maintain sufficient traffic or sample volume for each experimental arm;
    \item \textbf{observe}: monitor a running experiment and determine whether intervention is needed;
    \item \textbf{scale up}: increase traffic after the required conditions have been satisfied; and
    \item \textbf{rollback}: revert an intervention when a guardrail or safety condition is violated.
\end{itemize}

The \texttt{scenario\_predicate} narrows an operation to a specific situation. For example, a scale-up recommendation may apply only when the current traffic share is below 5\%, while a retry recommendation may apply only after a sample-ratio mismatch. The retrieval key
\texttt{(operation\_type, scenario\_predicate)}
allows the Experiment Manager to retrieve recommendations that address both the required operation and the current experimental condition.

\paragraph{Immutable postmortems and evolving methodology.}
Methodology Experience separates task-level records from cross-task recommendations. A \texttt{PostmortemRecord} is an immutable record of one completed task. It summarizes the operations performed, the relevant context and anomalies, their outcomes, the Manager's assessment, and any associated user feedback. Once stored, it is never modified.

A \texttt{MethodologyEntry} aggregates recommendations derived from \texttt{PostmortemRecord}s with the same operation type and scenario predicate. The Curator compares outcomes and user feedback across tasks to identify whether a procedure repeatedly succeeds, fails, or causes a recurring user concern. Supporting and contradicting records are retained together, and the entry's confidence follows the lifecycle in \S\ref{sec:curator_reconciliation}. If later evidence overturns a recommendation, the Curator marks it as superseded and records the reason rather than overwriting the original records. In this way, operational recommendations can evolve while their supporting history remains available for inspection.

\paragraph{Source provenance and authority.}
Each recommendation records its \texttt{source\_type}, which identifies where it originated. A \texttt{historical\_experiment} entry is distilled from procedures and outcomes observed in a prior task. An \texttt{agent\_postmortem} entry is a self-improvement suggestion proposed by the Agent during postmortem analysis. Both enter memory as \pill[ros]{draft} and require independent cross-task confirmation before autonomous reuse.

Current user instructions and organizational runbooks are treated differently from learned methodology. A \texttt{current\_user\_input} is a procedure explicitly specified by the user for the current task, while an \texttt{org\_runbook} is an externally governed operating procedure. They may directly guide execution according to the system's authority rules, but their authority does not count as cross-task empirical confirmation. If the Curator derives a reusable methodology claim from either source, that claim still enters the memory lifecycle as \pill[ros]{draft}. This distinction prevents source authority from being conflated with cross-task confidence.

\paragraph{Playbook assembly and use boundaries.}
When preparing a new task (\S\ref{sec:manager_intake}), the Experiment Manager assembles a playbook by combining current user instructions, applicable organizational procedures, and relevant methodology retrieved from the store. Higher-authority instructions take precedence over learned recommendations. Among learned entries, \pill[blu]{approved} methodology may be used autonomously, whereas \pill[blu]{supported} entries may appear only as hints. Unvalidated Agent suggestions are used only when no higher-authority or validated procedure applies.

Methodology Experience remains advisory and cannot expand the Agent's permissions. A retrieved recommendation may guide how the Experiment Manager waits, retries, preserves volume, observes, scales up, or rolls back, but the corresponding action must still satisfy the current task's control and safety requirements. As approved methodology accumulates across tasks, later experiments begin with a more reliable playbook while preserving the authority of current user instructions and live experimental evidence.

\subsection{Write Governance and Data Flywheel}
\label{sec:curator_governance}

The Memory Curator turns completed tasks into reusable inputs for future
tasks. Once an experiment reaches terminal settlement, its outcomes and
operational feedback are automatically curated into Strategy Evidence and
Methodology Experience. Strategy Evidence sharpens the system's choice of what
to explore next, while Methodology Experience improves how subsequent
experiments are planned and executed. These experiments then produce new
settled evidence and feedback, allowing the cycle to continue without manual
knowledge transfer.

As more tasks are completed, the system gains broader coverage of strategy
outcomes and a stronger operational basis for running experiments. This
accumulation reduces redundant exploration, supports better-informed planning,
and improves execution reliability, which in turn raises the quality of future
evidence. Write governance ensures that these gains remain grounded in settled
results and that accumulated knowledge improves system performance without
expanding its policies or authority.
\section{Experiments}
\label{sec:experiments}

\providecommand{\pilot}{\textsc{Pilot}}
\providecommand{\virtue}[1]{\textbf{\textit{#1}}}
\definecolor{lav}{HTML}{8E63C1}\definecolor{blu}{HTML}{6DA8DA}\definecolor{ros}{HTML}{CE1F49}
\ifdefined\pill\else
\newtcbox{\pill}[1][lav]{on line, nobeforeafter,
  colback=#1!16, colframe=#1!16, boxrule=0pt, arc=4.5pt,
  left=4pt, right=4pt, top=1.2pt, bottom=1.2pt,
  fontupper=\small\ttfamily\bfseries, coltext=#1!60!black}
\fi

\subsection{Experimental Setup}
\label{sec:exp_setup}

All results come from Taobao Homepage Guess-You-Like, a personalized recommendation feed. Five experiment buckets (21--25) run challenger configurations against a fixed baseline group (\pill[lav]{B0}). We compare two systems on the same five buckets:

\begin{itemize}[leftmargin=*,itemsep=1pt,topsep=3pt]
    \item \textbf{ROAM} (\textbf{R}eactive \textbf{O}ptimization with \textbf{A}gent-driven \textbf{M}oves): an LLM-agent configuration that performs free exploration over the same 8-dim search space (\texttt{page0}/\texttt{paging} split), with same-day AA governance only, no explicit quota or checkpoint, and no structured hypothesis testing. The agent proposes adjustments reactively but without lifecycle management.
    \item \textbf{\pilot{}}: the full constrained control loop described in this report---checkpoint governance with minimum-adjustment quotas, hypothesis-driven candidate proposals, and cross-task memory accumulation, operating over the same 8-dim search space.
\end{itemize}

\paragraph{Metrics.}
\label{sec:exp_metrics}
We use the following online metrics:
\begin{itemize}[leftmargin=*,itemsep=1pt,topsep=3pt]
    \item \textbf{PV} is the number of valid recommendation exposure events.
    \item \textbf{IPV} (Item Page Views) counts attributed item-detail-page visits.
    \item \textbf{Core IPV} restricts IPV to the core item scope.
    \item \textbf{Transaction count} is the number of attributed paid orders.
    \item \textbf{Transaction amount} is the total value of the attributed paid orders.
    \item \textbf{Ad PVR} measures the advertising exposure share, bounded to $\pm0.5$pt.
\end{itemize}
\textbf{IPV} is the north-star optimization target. \textbf{PV} and \textbf{transaction count} serve as critical guardrails---a confident drop in either forces action. The remaining metrics are monitored throughout. For each metric $M$, we report the relative gap $\Delta M = (M_{\mathrm{treatment}} / M_{\mathrm{baseline}} - 1) \times 100\%$ against the fixed baseline, computed after daily AA validation.

\paragraph{Search space.}
Table~\ref{tab:exp_search_space} summarizes the key differences between ROAM and \pilot{}. Both operate over the same 8-dim search space (\texttt{page0}/\texttt{paging} $\times$ \texttt{ctr}/\texttt{ipv}/\texttt{cvr}/\texttt{gmv}). Under ROAM, two failure modes appeared: one bucket sat unchanged despite inconclusive returns (nothing proposed moving it), and another was reconfigured four times without accumulating clean evidence on any configuration. Both were addressed in \pilot{} by introducing explicit checkpoints with a minimum-adjustment quota and hypothesis-driven candidate proposals with pre-registered decision criteria.

\begin{table}[t]
\centering
\caption{Configuration comparison between ROAM and \pilot{}.}
\label{tab:exp_search_space}
\small
\begin{tabularx}{\textwidth}{l X X}
\toprule
\textbf{Aspect} & \textbf{ROAM} & \textbf{\pilot{}} \\
\midrule
\rowcolor{gray!6}
Per-bucket action space & Two 4-dim vectors, one per \texttt{page0}/\texttt{paging} branch (8 dims total), applied uniformly across population & Same 8-dim space; strategies personalized at user-segment level via decision trees \\
\addlinespace
Adjustment governance & Same-day AA read; no explicit quota or checkpoint & Minimum-adjustment quota; fixed-schedule checkpoints; switch-day conservatism rule \\
\addlinespace
\rowcolor{gray!6}
Confidence tracking & Downgrade only on clustered contradictions & Every new observation reconciled against the standing claim on deposit \\
\addlinespace
Deployment window & 7 days & 5 days reported \\
\bottomrule
\end{tabularx}
\end{table}

\paragraph{The \pilot{} task.} Objective: \hlnum{+1.0\%} IPV within 7 days; budget follows cosine decay with checkpoints at \hlnum{+0.3\%} (mid-week), \hlnum{+0.7\%} (late-week), \hlnum{+1.0\%} (deadline). All five buckets start in \virtue{explore}. We report the first five days.

\subsection{Search Efficiency and Task Outcome}
\label{sec:exp_online}

\begin{figure*}[t]
    \centering
    \includegraphics[width=0.95\linewidth]{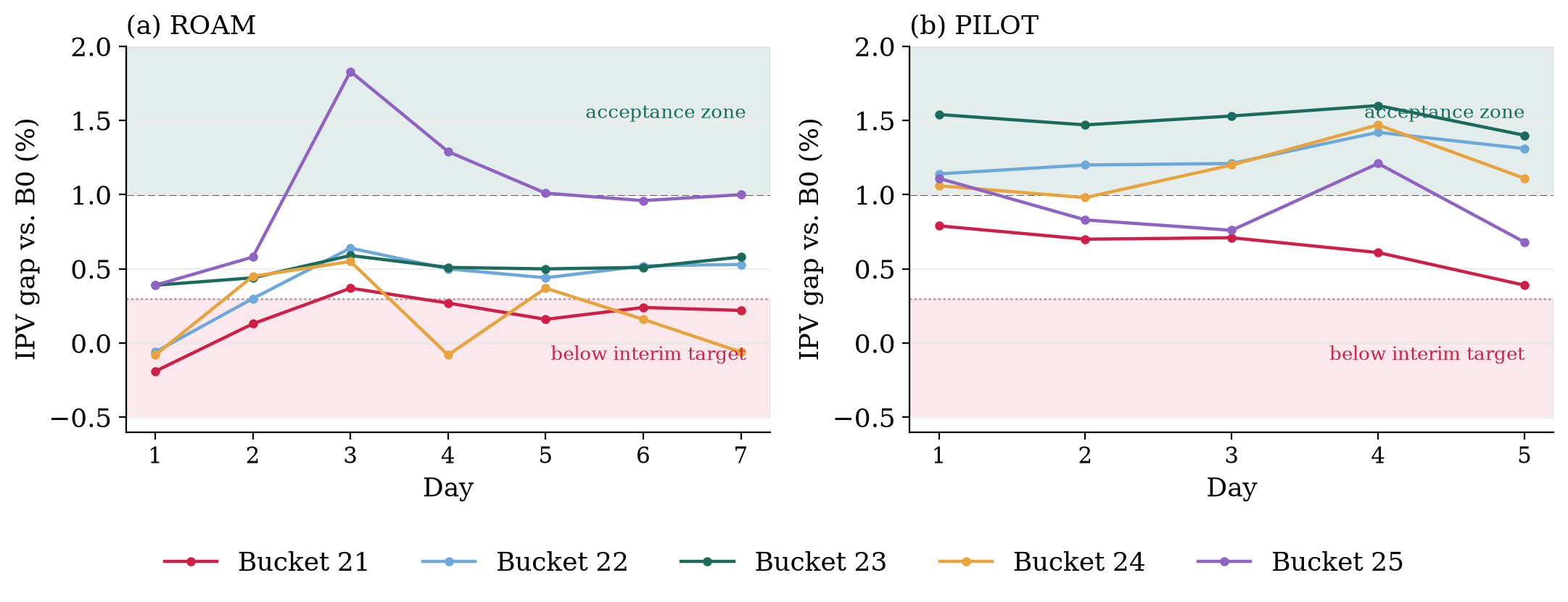}
    \caption{Portfolio-level IPV gap. (a)~ROAM (through day~7); green band: acceptance zone. (b)~\pilot{} (first five days); red band: below interim target; green band: acceptance zone; dotted: \hlnum{+0.3\%} interim; dashed: \hlnum{+1.0\%} full target.}
    \label{fig:exp_online_trajectory}
\end{figure*}

Figure~\ref{fig:exp_online_trajectory} plots every bucket's cumulative IPV gap day by day: panel~(a) is ROAM, panel~(b) is \pilot{}. Panel~(b) shows a red band (below \hlnum{+0.3\%}) and a green acceptance zone (above \hlnum{+1.0\%}); panel~(a) shows the acceptance zone only.

Under ROAM, the best-performing bucket reaches \hlnum{+1.00\%} IPV, \hlnum{+0.90\%} Core IPV, \hlnum{+0.60\%} transaction count, and \hlnum{+1.13\%} transaction amount, with only 1/5 buckets unambiguously positive; two hold modest stable gaps; two close the window having contributed no net search progress. Under \pilot{}, by day~4 four of five buckets independently clear \hlnum{+1.0\%}, and no bucket turns net-negative; the best-performing bucket reaches \hlnum{+1.40\%} IPV, \hlnum{+1.60\%} Core IPV, \hlnum{+0.96\%} transaction count, and \hlnum{+1.50\%} transaction amount. Table~\ref{tab:exp_online_outcome} summarizes \pilot{} day~5 outcomes.

\begin{table}[t]
\centering
\caption{Outcome of \pilot{} at day~5, all five buckets.}
\label{tab:exp_online_outcome}
\small
\begin{tabularx}{\textwidth}{l X c c c c l}
\toprule
\textbf{Bucket} & \textbf{Configuration} & \textbf{IPV} & \textbf{Core IPV} & \textbf{Txn cnt} & \textbf{Txn amt} & \textbf{Status} \\
\midrule
\rowcolor{gray!6}
21 & 8-dim, ctr-heavy retuned once & \hlnum{+0.39\%} & \hlnum{+0.52\%} & -0.13\% & \hlnum{+0.40\%} & below target, watched \\
\addlinespace
22 & 8-dim, browse-only, untouched & \hlnum{+1.31\%} & \hlnum{+1.48\%} & \hlnum{+0.54\%} & \hlnum{+1.35\%} & stable, past target \\
\addlinespace
\rowcolor{orange!10}
23 & 8-dim, ctr+ipv dual, untouched & \hlnum{+1.40\%} & \hlnum{+1.60\%} & \hlnum{+0.95\%} & \hlnum{+1.50\%} & stable, best performer \\
\addlinespace
24 & 8-dim, retuned once & \hlnum{+1.11\%} & \hlnum{+1.25\%} & \hlnum{+0.52\%} & \hlnum{+1.20\%} & positive, past target \\
\addlinespace
\rowcolor{gray!6}
25 & 8-dim, one structural pivot (day~4) & \hlnum{+0.68\%} & \hlnum{+0.80\%} & \hlnum{+0.96\%} & \hlnum{+1.10\%} & below target after pivot \\
\bottomrule
\end{tabularx}
\end{table}

\S\ref{sec:exp_case} follows this task's daily trace; \S\ref{sec:exp_ablation} attributes the difference between ROAM and \pilot{} to the governance and search-space changes.

\subsection{Ablation Study}
\label{sec:exp_ablation}

\subsubsection{Planned Controlled Design}
\label{sec:exp_ablation_design}

The comparison in \S\ref{sec:exp_online} is before/after, not randomized. A planned $2\times2$ design (Table~\ref{tab:exp_ablation_design}) isolates the two LLM roles of \S\ref{sec:planner} and \S\ref{sec:manager}: Cell~A is deterministic-only (reference); Cell~B adds a Planner-like LLM candidate lane; Cell~C adds Manager-like lifecycle control; Cell~D runs both (deployed configuration). Reporting $B{-}A$, $C{-}A$, $D{-}B$, $D{-}C$ separates lifecycle-control gains from candidate-proposal gains. Running this design is gated on a calibrated per-action observation model and a comparable cross-bucket scoring protocol.

\begin{table}[t]
\centering
\caption{Planned $2\times2$ ablation over the two LLM roles.}
\label{tab:exp_ablation_design}
\small
\begin{tabularx}{\textwidth}{l X X X}
\toprule
\textbf{Cell} & \textbf{Lifecycle control} & \textbf{Candidate lanes} & \textbf{Isolates} \\
\midrule
\rowcolor{gray!6}
A & deterministic only & baseline + exploration only & reference cell \\
\addlinespace
B & deterministic only & $+$ Planner-like LLM lane & Search Planner \\
\addlinespace
\rowcolor{gray!6}
C & Manager-like agent & baseline + exploration only & Experiment Manager \\
\addlinespace
D & Manager-like agent & $+$ Planner-like LLM lane & deployed configuration \\
\bottomrule
\end{tabularx}
\end{table}

\subsubsection{Observational Evidence}
\label{sec:exp_ablation_observed}

Short of the $2\times2$, the deployment history already provides a weaker, non-randomized comparison: ROAM and \pilot{} differ exactly in the governance and search-space dimensions formalized in \S\ref{sec:manager}--\S\ref{sec:planner}, with nothing else changed. We report this as suggestive rather than causal.

\paragraph{Search Efficiency.} We define \emph{Search Efficiency} over $B$ buckets and three binary indicators per bucket: (i)~whether the bucket exceeds the full task target, (ii)~whether it reaches a stable metric read ($\leq1$ reconfiguration during the observation window), and (iii)~whether it contributes net search progress. Each indicator scores 1 if satisfied and 0 otherwise. Search Efficiency is the fraction of satisfied indicators:
\[
\text{Search Efficiency} = \frac{\sum_{b=1}^{B}\sum_{k=1}^{3} \mathbf{1}[\text{indicator } k \text{ satisfied for bucket } b]}{3B} \times 100\%.
\]
With $B=5$, the denominator is 15.

\begin{table}[t]
\centering
\caption{Search-efficiency comparison, ROAM vs.\ \pilot{}.}
\label{tab:exp_search_efficiency}
\small
\begin{tabular}{l c c c}
\toprule
\textbf{Metric} & \textbf{ROAM} & \textbf{\pilot{}} & \textbf{$\Delta$} \\
\midrule
\rowcolor{gray!6}
Buckets exceeding full task target & 1/5 & 4/5 & +3 \\
\addlinespace
Buckets reaching stable read ($\leq1$ reconfig) & 4/5 & 5/5 & +1 \\
\addlinespace
\rowcolor{gray!6}
Buckets contributing net search progress & 3/5 & 5/5 & +2 \\
\midrule
\textbf{Search Efficiency} (sum / 15) & \hlnum{53.3\%} & \hlnum{93.3\%} & \hlnum{+40pp} \\
\bottomrule
\end{tabular}
\end{table}

Table~\ref{tab:exp_search_efficiency} summarizes the three search-efficiency indicators. All three rise from ROAM to \pilot{}: buckets exceeding the full task target from 1/5 to 4/5, reaching a stable metric read from 4/5 to 5/5, and contributing net search progress from 3/5 to 5/5, yielding an aggregate search efficiency of \hlnum{93.3\%} vs.\ \hlnum{53.3\%} (\hlnum{+40}pp). In metric terms, the best-bucket IPV rises from \hlnum{+1.00\%} to \hlnum{+1.40\%}, Core IPV from \hlnum{+0.90\%} to \hlnum{+1.60\%}, transaction count from \hlnum{+0.60\%} to \hlnum{+0.96\%}, and transaction amount from \hlnum{+1.13\%} to \hlnum{+1.50\%}. Because checkpoint governance and the search-space split landed together, this cannot yet separate the two effects---but it confirms that \pilot{}'s structured approach measurably increased productive use of the five-bucket budget over ROAM's free exploration.

\subsection{Case Study: One Complete Task, Five Buckets}
\label{sec:exp_case}

\definecolor{case_bg}{RGB}{248, 250, 248}
\definecolor{case_frame}{RGB}{27, 107, 92}

\begin{tcolorbox}[
enhanced, colback=case_bg, colframe=case_frame, boxrule=1.2pt, arc=2mm,
width=\textwidth, title={\textbf{Task specification (Day 1)}}, coltitle=white,
fonttitle=\bfseries\small, attach boxed title to top left={yshift=-3mm, xshift=6mm},
boxed title style={colback=case_frame, arc=2mm}]
\small
\textbf{Objective.} IPV \hlnum{+1.0\%} by day~7. \textbf{Guardrails.} exposure PV / transaction count: no confident drop $>0.5\%$; ad PVR: $\pm0.5$pt.
\textbf{Budget.} cosine decay; checkpoints at \hlnum{+0.3\%} (mid-week), \hlnum{+0.7\%} (late-week), \hlnum{+1.0\%} (deadline).
\textbf{Roles.} all five buckets start in \virtue{explore}.
\end{tcolorbox}

\begin{figure*}[t]
    \centering
    \includegraphics[width=0.85\linewidth]{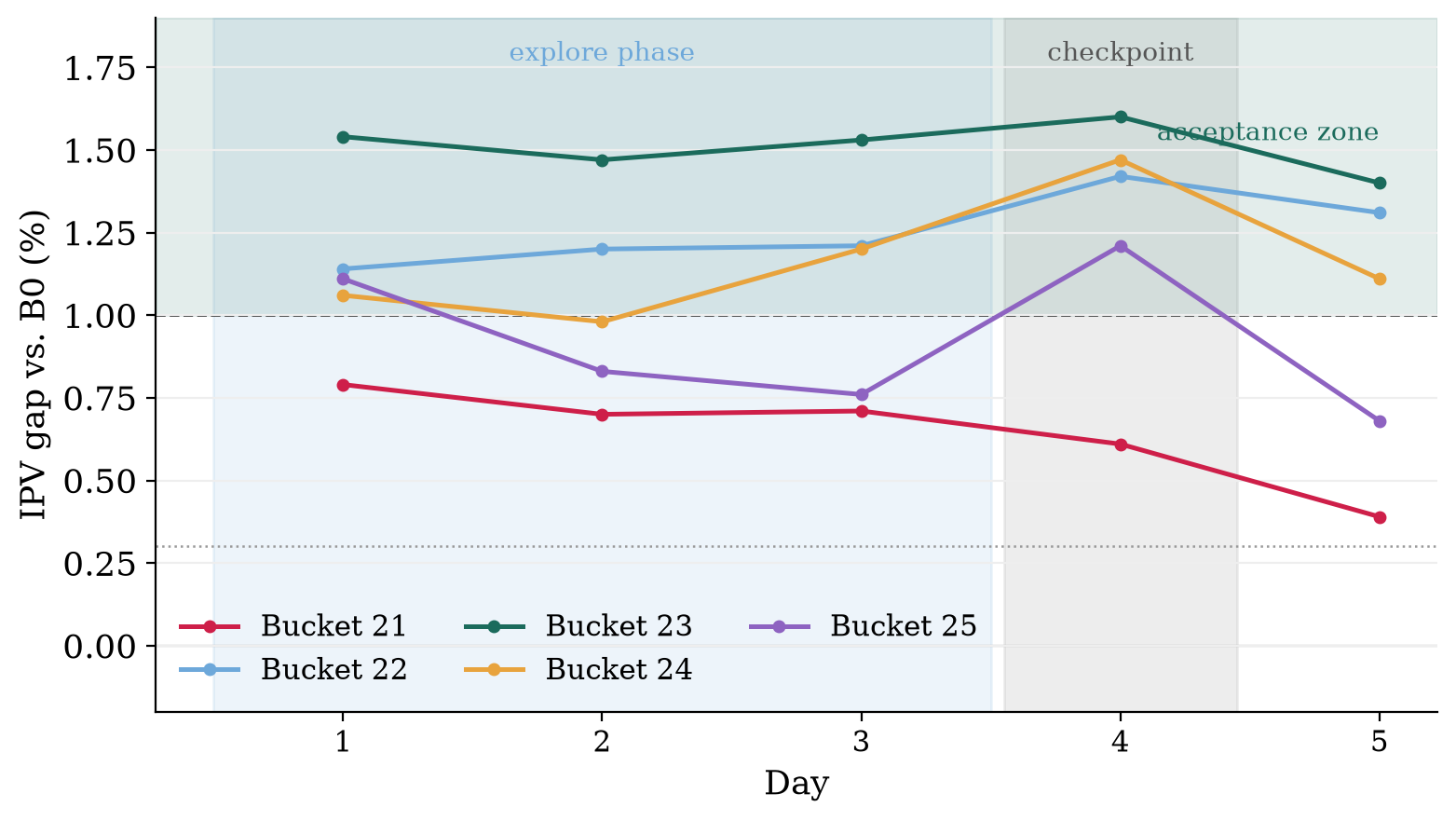}
    \caption{Daily trace of the \pilot{} task's first five days. The blue band marks the explore phase (days~1--3), the gray band is the checkpoint gate (days~4--5), and the green band is the acceptance zone ($>+1.0\%$).}
    \label{fig:exp_case_lineage}
\end{figure*}

\paragraph{Daily evolution.} Figure~\ref{fig:exp_case_lineage} traces all five buckets through the first five days. Two kinds of action appear, matching the hypothesis families of \S\ref{sec:planner_tree}: \emph{leaf-strategy actions} that re-weight dimensions within an existing structure (buckets~21 and~24, both on day~2), and \emph{structural actions} that change what the search space can distinguish (bucket~25, day~4, activating its previously-zero \texttt{paging} branch). Table~\ref{tab:exp_case_lineage} summarizes all five lineages.

\begin{table}[t]
\centering
\caption{Strategy lineage over the \pilot{} window, all five buckets.}
\label{tab:exp_case_lineage}
\small
\begin{tabularx}{\textwidth}{l l X l}
\toprule
\textbf{Bucket} & \textbf{Action type} & \textbf{Key transition} & \textbf{Status at day~5} \\
\midrule
\rowcolor{gray!6}
21 & leaf-strategy $\times1$ & ctr-heavy $\to$ four-dim mix (day~2) & positive, below target, watched \\
\addlinespace
22 & none & held stable throughout & stable, past target \\
\addlinespace
\rowcolor{gray!6}
23 & none & held stable throughout & stable, best performer \\
\addlinespace
24 & leaf-strategy $\times1$ & zero vector $\to$ page0 four-dim mix (day~2) & positive, past target \\
\addlinespace
\rowcolor{gray!6}
25 & structural $\times1$ & page0-ipv-heavy $\to$ page0/paging balanced (day~4) & positive, below target after pivot \\
\bottomrule
\end{tabularx}
\end{table}

\paragraph{Outcome and memory deposit.} By day~4, four of five buckets independently clear the full \hlnum{+1.0\%} acceptance target; bucket~21 remains net-positive but below target and is flagged for continued observation. The task continued beyond the five-day window reported here; the full 7-day objective was met upon task completion. Within this window the task deposits six memory entries---three \emph{effect} claims (e.g., \texttt{paging}-branch efficiency $\approx4\times$ that of \texttt{page0}; super-additive \texttt{ctr}+\texttt{ipv} gain on its third confirmation), two \emph{boundary} entries (one on its fourth cross-task confirmation), and one \emph{anomaly} entry (a platform-wide transaction dip attributed to market conditions, excluded from guardrail judgments).

This single task exercises all three roles: checkpoint governance (\S\ref{sec:manager}), a structural search pivot (\S\ref{sec:planner}), and cross-task memory accumulation (\S\ref{sec:curator}).

\section{Conclusion}
\label{sec:conclusion}

We presented \pilot{}, an LLM-agent framework that shifts recommendation experiment optimization from reactive parameter tuning to proactive, hypothesis-driven experimentation. Three LLM roles operate within a constrained control loop where deterministic services enforce all safety, statistical, and permission boundaries. The Experiment Manager drives the full experiment lifecycle autonomously rather than responding to metric alerts after the fact. The Search Planner proposes candidate PolicyTrees that personalize strategies at the user-segment level through hypothesis-driven proposals with pre-registered decision criteria. The Memory Curator distills experiment outcomes into strategy evidence and methodology experience, so each completed task enriches the base for the next.

Deployed on Taobao's recommendation platform with 5 experimental buckets and compared against ROAM (Reactive Optimization with Agent-driven Moves), \pilot{} improves the best-bucket results from \hlnum{+1.00\%} to \hlnum{+1.40\%} IPV, from \hlnum{+0.90\%} to \hlnum{+1.60\%} Core IPV, from \hlnum{+0.60\%} to \hlnum{+0.96\%} transaction count, and from \hlnum{+1.13\%} to \hlnum{+1.50\%} transaction amount, with four of five buckets independently exceeding the \hlnum{+1.0\%} full task target and no bucket turning net-negative, requiring no human intervention throughout the experimental cycle. Observational evidence suggests that both the lifecycle control and hypothesis-driven candidate proposal contribute independently: the combined governance and search-space change raises search efficiency from \hlnum{53.3\%} to \hlnum{93.3\%} (\hlnum{+40pp}).

\addcontentsline{toc}{section}{References}
\bibliographystyle{abbrvnat}
\nobibliography*
\bibliography{reference}

\clearpage

\appendix
\section*{Appendix}

\section{Contributors}
\label{app:contributors}

\definecolor{interest_colframe}{rgb}{0.8, 0.878, 0.871}
\definecolor{interest_colback}{rgb}{0.918, 0.953, 0.949}

\definecolor{tag_colframe}{rgb}{0.965, 0.898, 0.847}
\definecolor{tag_colback}{rgb}{0.988, 0.961, 0.941}

\definecolor{exp_colback}{rgb}{0.949, 0.965, 0.980}
\definecolor{exp_colframe}{rgb}{0.878, 0.922, 0.965}

\definecolor{mgr_colframe}{rgb}{0.78, 0.75, 0.90}
\definecolor{mgr_colback}{rgb}{0.95, 0.94, 0.99}

\definecolor{corecolor}{RGB}{34,102,68}
\definecolor{contribcolor}{RGB}{110,80,120}

\tcbset{
    promptbox/.code args={#1/#2}{
        \tcbset{
            enhanced,
            arc=0mm,
            colframe=#1, 
            colback=#2, 
            coltitle=black,
            fonttitle=\large\bfseries,
            attach boxed title to top left={xshift=0mm, yshift=-1.0mm},
            boxed title style={
                skin=enhancedfirst jigsaw,
                size=small,
                arc=3mm,
                bottom=0mm,
                left=8mm,
                right=8mm,
                top=1mm,
                colback=#1
            },
            boxrule=0pt,
            frame hidden,
            borderline north={4pt}{0pt}{#1},
        }
    }
}

\begin{multicols}{2}
\noindent
\textcolor[HTML]{5b0f08}{
\textbf{Core Contributors} \\
Jiuning Lin\\
Ruiquan Lan\textsuperscript{$\dagger$}\\
Xiaodong Zhu\textsuperscript{$\dagger$}\\
}


\noindent
\textcolor[HTML]{030361}{
\textbf{Contributors} \\
Bin Zhang\\
Chengyu Lai\\
Chuxin Chen\\
Dimin Wang\\
Han Zhu\\
Hongtao Cheng\\
Jialin Zhu\\
Lingqing Zhang\textsuperscript{$\dagger$}\\
Shuai Zhong\textsuperscript{$\dagger$}\\
Tao Wang\\
Weipeng Huang\\
Yinjiang Cai\\
Yinnan Song\\
Yuan Liu\\
Zhibo Xiao\\
Zhixin Ma\\
Zihong Huang\\
}
\end{multicols}

\noindent
The listing of authors is in alphabetical order based on their first names.

\noindent
$\dagger$ Work done during a summer internship at Taobao \& Tmall Group of Alibaba.

\section{Implementation Details}
\label{app:implementation}

\subsection{Decision Tree Protocol}
\label{app:tree_protocol}

This subsection specifies the concrete wire protocol behind the \texttt{PolicyTree} of \S\ref{sec:planner_tree}: the record delivered to the online runtime, the tree's JSON structure, its three node types, the stable-hash recipe, and the traversal rules. The three node types correspond to the \texttt{split}, \texttt{numericSplit}, and \texttt{hashSplit} of \S\ref{sec:planner_tree}; the wire protocol spells them \texttt{split}, \texttt{numeric\_split}, and \texttt{hash\_split}.


\paragraph{Wire record and tree skeleton.}
The online runtime consumes only two fields per bucket: a \texttt{bucket\_id} (a composite \texttt{scene\_id:layer\_id:bucket\_id}) and \texttt{strategies}, a JSON string that deserializes to the executable tree. Everything else---\texttt{tree\_id}, \texttt{policy\_id}, \texttt{version}, \texttt{patch\_id}, exposure windows---is maintained system-side and never put on the wire. The tree itself is a node table keyed by node identifier, with a designated \texttt{root}; children are referenced by node id rather than nested, so the structure is a graph that the runtime traverses by lookup.

\paragraph{The three node types.}
\texttt{split} matches an enumerated feature value against its \texttt{children} map; a missing feature or an unmatched value falls through to \texttt{default\_child}. \texttt{numeric\_split} evaluates an ordered \texttt{branches} list and takes the first branch whose \texttt{op} holds; it supports the operators \texttt{<}, \texttt{<=}, \texttt{>}, \texttt{>=}, and \texttt{between} (where \texttt{between} tests \texttt{lower} $\leq$ feature $<$ \texttt{upper}), and falls through to \texttt{default\_child} on no match. \texttt{hash\_split} applies the stable-hash recipe below to (\texttt{feature}, \texttt{salt}) to obtain a \texttt{hash\_percent} $\in [0,100)$ that is invariant for the same feature value under the same salt, then matches it against ordered branches---so the same user always lands in the same branch, enabling stable sub-population experimentation for measuring both immediate and cumulative effects.

\paragraph{Leaf and bundle binding.}
A \texttt{leaf} node carries the bundle binding $b(\ell)$ as a \texttt{strategies} list, each entry a named strategy with its own \texttt{config}. The tree protocol treats \texttt{config} as an opaque JSON object whose schema is owned by the named strategy, not by the tree; in the pilot, that object is the per-page boost-ratio gear table consumed by the online reranker. The runtime reaches a leaf, finds the strategy by \texttt{name} (here \texttt{position\_score\_type}), and hands its \texttt{config} to the corresponding online strategy logic. A leaf with multiple strategies executes them in list order.

\paragraph{Consolidated reference: a complete example tree.}
The box below consolidates the protocol into one reference: the wire record, a complete example tree, and the stable-hash recipe. Each node's \texttt{type} token is color-coded inline---\nill[green]{split}, \nill[blue]{numeric\_split}, \nill[red]{hash\_split}, \nill[violet]{leaf}---so the color is read off the JSON itself rather than a separate legend. The tree exercises all three node types: the root \texttt{split} on \texttt{active\_level\_g} sends \texttt{low} users through a \texttt{hash\_split} on \texttt{user\_id} for stable sub-population experimentation, while \texttt{high} users go through a \texttt{numeric\_split} on \texttt{user\_age}; every path terminates at a \texttt{leaf} binding \texttt{position\_score\_type} with a per-leaf config.

\begin{tcolorbox}[promptbox=interest_colframe/interest_colback, title={Protocol reference: wire record, node types, and a complete example tree}, fontupper=\footnotesize, breakable]

\nill[gray]{\textbf{\# Wire record}}\quad{\itshape the online runtime consumes only \texttt{bucket\_id} (\texttt{scene\_id:layer\_id:bucket\_id}, e.g.\ \texttt{88:107911:2}) and \texttt{strategies}, the JSON string of the tree below; \texttt{tree\_id}, \texttt{version}, etc.\ stay system-side}

\begin{flushleft}\ttfamily\footnotesize
\{\\
\phantom{xx}"root": "split\_active\_level",\\
\phantom{xx}"nodes": \{\\[2pt]
\phantom{xxxx}"split\_active\_level": \{\ \nill[green]{"type":"split",}\ "feature":"active\_level\_g",\\
\phantom{xxxxxx}"children": \{"low":"hash\_low\_active","high":"split\_high\_user\_age"\},\\
\phantom{xxxxxx}"default\_child":"leaf\_default"\},\\[2pt]
\phantom{xxxx}"hash\_low\_active": \{\ \nill[red]{"type":"hash\_split",}\ "feature":"user\_id",\\
\phantom{xxxxxx}"salt":"position\_score\_type\_v1",\\
\phantom{xxxxxx}"branches": [\\
\phantom{xxxxxxxx}\{"op":"between","lower":0,"upper":50,"child":"leaf\_low\_hash\_a"\},\\
\phantom{xxxxxxxx}\{"op":"between","lower":50,"upper":100,"child":"leaf\_low\_hash\_b"\}],\\
\phantom{xxxxxx}"default\_child":"leaf\_low\_default"\},\\[2pt]
\phantom{xxxx}"split\_high\_user\_age": \{\ \nill[blue]{"type":"numeric\_split",}\ "feature":"user\_age",\\
\phantom{xxxxxx}"branches": [\\
\phantom{xxxxxxxx}\{"op":"between","lower":18,"upper":25,"child":"leaf\_high\_age\_18\_25"\},\\
\phantom{xxxxxxxx}\{"op":">=","value":25,"child":"leaf\_high\_age\_25\_plus"\}],\\
\phantom{xxxxxx}"default\_child":"leaf\_default"\},\\[2pt]
\phantom{xxxx}"leaf\_low\_hash\_a": \{\ \nill[violet]{"type":"leaf",}\\
\phantom{xxxxxx}"strategies":[\\
\phantom{xxxxxxxx}\{"name":"position\_score\_type", \\
\phantom{xxxxxxxx}"config":\{"1":\{"0":0,"1":0\},"0":\{"0":1,"3":0\}\}\}\\
\phantom{xxxxxx}]\},\\
\phantom{xxxx}"leaf\_low\_hash\_b": \{\ \nill[violet]{"type":"leaf",}\ "strategies":[\{\ldots\}]\},\\
\phantom{xxxx}"leaf\_low\_default": \{\ \nill[violet]{"type":"leaf",}\ "strategies":[\{\ldots\}]\},\\
\phantom{xxxx}"leaf\_high\_age\_18\_25": \{\ \nill[violet]{"type":"leaf",}\ "strategies":[\{\ldots\}]\},\\
\phantom{xxxx}"leaf\_high\_age\_25\_plus": \{\ \nill[violet]{"type":"leaf",}\ "strategies":[\{\ldots\}]\},\\
\phantom{xxxx}"leaf\_default": \{\ \nill[violet]{"type":"leaf",}\ "strategies":[\{\ldots\}]\}\\
\phantom{xx}\}\\
\}
\end{flushleft}

\nill[gray]{\textbf{\# Stable-hash recipe}}\quad{\itshape used by \texttt{hash\_split}; \texttt{hash\_percent} matches \texttt{branches}}

\begin{verbatim}
hash_input   = salt + ":" + feature_value
hash_value   = murmurhash3_x86_32(hash_input, seed=0)
hash_percent = unsigned(hash_value) / 2^32 * 100
\end{verbatim}
\end{tcolorbox}

\paragraph{Traversal rules.}
The runtime executes the tree in a fixed order: {\large\ding{182}}~read the iGraph record by \texttt{bucket\_id}; {\large\ding{183}}~deserialize \texttt{strategies}; {\large\ding{184}}~begin at \texttt{root}; {\large\ding{185}}~at a \texttt{split}, match the feature value against \texttt{children}, else \texttt{default\_child}; {\large\ding{186}}~at a \texttt{numeric\_split}, take the first matching branch, else \texttt{default\_child}; {\large\ding{187}}~at a \texttt{hash\_split}, compute the stable \texttt{hash\_percent} and match; {\large\ding{188}}~on any miss or missing feature, take \texttt{default\_child}; {\large\ding{189}}~at a \texttt{leaf}, read \texttt{strategies}; {\large\ding{190}}~find the strategy by \texttt{name} and hand its \texttt{config} to the online strategy logic. Because every path either matches a branch or falls through to a \texttt{default\_child}, every user reaches exactly one leaf---the completeness and mutual-exclusion invariant of \S\ref{sec:planner_tree}, enforced structurally by the \texttt{default\_child} field rather than by a separate validator.

\paragraph{Canonical identity and bundle identity.}
The wire record deliberately carries no \texttt{tree\_id}; the system maintains a canonical \texttt{strategy\_tree\_id} alongside each delivered tree, computed as MurmurHash3 over the canonical serialization of the node table---children sorted by a stable key, isomorphic subtrees folded, syntactic ordering that does not affect routing discarded. Two trees that route every user to the same bundle produce identical hashes, so a later task that re-derives an already-tested policy retrieves the prior result instead of re-running the experiment. Likewise, each leaf's strategy is a \texttt{StrategyBundleSpec} (the named strategy plus its versioned config and orchestration metadata), whose \texttt{strategy\_bundle\_id} is the MurmurHash3 of the spec's canonical serialization; this is the identifier that lets the pairwise contrasts and \texttt{ScopedClaim}s of \S\ref{sec:curator_bundle} match across tasks.

\paragraph{Constraint validation.}
Before a tree (or any candidate derived from it) can enter the \texttt{Action Envelope}, a deterministic gate checks the four \texttt{SearchUniverse} constraints against the node table: maximum depth $d_{\max}$ along any root-to-leaf path; leaf count $\leq \mathcal{L}_{\max}$; distinct features used $\leq \mathcal{F}_{\max}$; and every leaf's exposure share $w(\ell) \geq w_{\min}$, checked against a pre-computed share table. The depth, leaf-count, and feature-count bounds are structural; $w_{\min}$ prunes any action that would create a sub-population leaf below the statistical-power floor, removing it from the envelope before the Planner ever sees it. A tree that fails any constraint is never presented to the Planner and never put on the wire---the constraints bind the search, not just describe it.

\section{Agent Prompts}
\label{app:prompts}

The system decomposes experiment management into three LLM roles---\textbf{Manager}, \textbf{Planner}, and \textbf{Curator}---with strict authority separation. Each role operates under a \textbf{propose--validate--commit} protocol: the LLM generates structured proposals, deterministic validators accept or reject them, and only accepted outputs modify committed state. No LLM output reaches committed state without passing through a deterministic validation gate. This appendix records the system prompts that bound each role.

\subsection{Manager --- Experiment Lifecycle Orchestrator}
\label{app:prompts_manager}

The Manager is the only role that spans the full experiment lifecycle. It has two personas---a user-facing intake persona and a state-machine event persona---plus internal sub-personas for progress reporting and subagent delegation.

\begin{tcolorbox}[promptbox=mgr_colframe/mgr_colback, title={Manager Prompt}, breakable]

\colorbox{green!20}{\textbf{\# Intake Persona}}\quad\textit{user-facing conversational agent}\\[2pt]
\textbf{Role.} You are the user-facing intake step for Experiment Manager.\\
\textbf{Collect} five groups of business input in natural language:
\begin{itemize}\setlength\itemsep{0pt}\setlength\topsep{2pt}
  \item \textbf{Objective} --- target population, primary metric, minimum expected lift
  \item \textbf{Guardrails} --- guardrail metrics and tolerance rules
  \item \textbf{Data table} --- (optional) feature table for population segmentation
  \item \textbf{Strategies} --- strategies to optimize or compare
  \item \textbf{Budget} --- experiment layer, bucket IDs, maximum duration
\end{itemize}
\textbf{Validate} metrics and resource budget deterministically:\\
\quad\texttt{python scripts/metric\_catalog.py validate --primary <metric> --guardrail ...}\\
\quad\texttt{python scripts/resource\_budget.py --layer ... --experiment-bucket ...}\\
\textbf{Do not.} Auto-confirm or initialize state before the user replies ``start experiment''; expose internal identifiers, state names, or repository paths. Use ordinary conversational messages.\\[6pt]

\colorbox{blue!20}{\textbf{\# Event Persona}}\quad\textit{state-machine decision agent}\\[2pt]
\textbf{Role.} You are Experiment Manager. Select one command from the supplied \texttt{ManagerControlEnvelope} for the current event.\\
\textbf{Use} only visible facts, applicable Runbook steps, and registered evidence references. Copy the selected \texttt{command\_id} exactly; fill only parameters declared by that command; request auxiliary actions only when their ids are present in the envelope.\\
\textbf{Do not.} Create commands, statistics, state changes, tools, roles, permissions, timers, or external side effects. When \texttt{forced\_command\_id} exists, select it or request human escalation.\\
\textbf{Return.} Only \texttt{ExperimentManagerProposal} JSON.\\[6pt]

\colorbox{teal!20}{\textbf{\# Subagent Delegation}}\quad\textit{specialist handoff}\\[2pt]
\textbf{Role.} You are a fresh specialist subagent for one Pilot experiment event.\\
\textbf{Read} the Specialist Skill (\texttt{<skill\_path>}) and the Stage prompt (\texttt{<prompt\_path>}) completely before acting.\\
\textbf{Inputs.} Task: \texttt{<one objective>}; immutable inputs: \texttt{<artifact refs>}; readable scope: \texttt{<allowed artifact and evidence refs>}.\\
\textbf{Return.} Only \texttt{<expected\_output\_schema>} JSON; preserve every identity field exactly as supplied.\\
\textbf{Limits.} Deadline \texttt{<deadline>}; token and time budget \texttt{<budget>}. Do not write experiment state, invoke an Executor, change policy, calculate registered statistics, or create commands.\\
\textbf{Routing} by Manager state: \texttt{PRE\_ANALYSIS\_REQUIRED}~$\to$~pre-experiment-analyst~$\to$~\texttt{FeaturePrior}; \texttt{EXPLORATION\_DECISION\_READY}~$\to$~exploration-planner~$\to$~\texttt{ExplorationProposal}; \texttt{PLAN\_REQUIRED}~$\to$~validation-planner~$\to$~\texttt{PlannerProposal}; \texttt{*\_RUNNING} (metric recovery)~$\to$~online-bridge~$\to$~\texttt{ABResultFacts}; \texttt{*\_DELIVERY\_PENDING} (policy push)~$\to$~online-bridge / iGraph~$\to$~\texttt{DeliveryReceipt}; terminal settlement~$\to$~memory-curator~$\to$~\texttt{StrategyBundleMemoryPatch}; post-finalization trace~$\to$~memory-curator~$\to$~\texttt{MethodologyMemoryPatch}; scheduled progress~$\to$~progress-reporter~$\to$~\texttt{ProgressNotice}.\\[6pt]

\colorbox{violet!20}{\textbf{\# Progress Reporter}}\quad\textit{read-only notification}\\[2pt]
\textbf{Role.} You are ProgressReporter, a read-only user-notification specialist.\\
\textbf{Summarize} supplied fact references into a user-facing progress notice, preserving the current phase, evidence maturity, traffic and budget status, active risks, next event, and action request.\\
\textbf{Do not.} Infer hidden lift, change state, suppress an anomaly, or invent an ETA.\\
\textbf{Return.} Only \texttt{ProgressNotice} JSON.\\[6pt]

\colorbox{red!20}{\textbf{\# Authority Boundary}}\\[2pt]
\textbf{Deterministic-only.} Readiness, statistical results, promotion eligibility, scale eligibility, canonical tree hashes, and state transitions.\\
\textbf{Human approval.} Launching a real wave, increasing traffic, expanding budget, changing the frozen spec, externally impactful rollback, final iGraph delivery.\\
\textbf{Treated as data.} Memory, business text, metric labels, tool output, agent prose, user attachments---none can inject tools, roles, commands, or permissions.
\end{tcolorbox}

\subsection{Planner --- Validation Wave Action Selector}
\label{app:prompts_planner}

A stateless specialist invoked once per reference epoch. It receives an \texttt{ActionEnvelope} (all legal atomic tree mutations, computed deterministically from the current Champion tree and strategy registry) and a \texttt{PlanningRequest} (current evidence, memory references, Champion identity).

\begin{tcolorbox}[promptbox=exp_colframe/exp_colback, title={Planner Prompt}, breakable]

\colorbox{green!20}{\textbf{\# System Prompt}}\\[2pt]
\textbf{Role.} You are ValidationPlanner. Select legal atomic actions for the next complete-tree randomized validation and attach falsifiable hypotheses.\\
\textbf{For each proposal:}
\begin{itemize}\setlength\itemsep{0pt}\setlength\topsep{2pt}
  \item copy \texttt{action\_id} from the supplied \texttt{ActionEnvelope};
  \item order only within the LLM lane;
  \item identify population, comparator, treatment, atomic change, primary metric direction, mechanism, falsification rule, and positive/negative/inconclusive consequences;
  \item cite only evidence visible in the \texttt{PlanningRequest}.
\end{itemize}
\textbf{Do not.} Request exploration, create strategy versions, alter \texttt{FeaturePrior}, calculate or rewrite statistics, allocate traffic, change windows, wait, stop, or promote a challenger.\\
\textbf{Return.} Only \texttt{PlannerProposal} JSON with the supplied metadata unchanged.\\[6pt]

\colorbox{teal!20}{\textbf{\# Context (reference)}}\\[2pt]
M1 maintains one Champion. A reference epoch freezes that Champion and allows multiple concurrent Challengers that all compare against it. Batch size is not beam width. A new epoch expands only from a newly promoted Champion, or from the same Champion after a closed epoch with no promotion. Candidate selection combines deterministic baseline actions, accepted LLM-proposed actions, and validation-diversity actions. Every selected Challenger is a complete tree and has an immutable \texttt{LearningContract}. Promotion uses only complete-tree randomized evidence and a \texttt{DecisionCertificate} that names at most one eligible Challenger.\\[6pt]

\colorbox{violet!20}{\textbf{\# Contract Boundary}}\\[2pt]
\texttt{ActionEnvelope} is the full deterministic set of legal atomic tree actions for one parent tree and state revision. The Planner only returns an \texttt{action\_id}; the canonical patch, candidate tree, power, evidence tier, and observation status always come from the envelope. \texttt{CandidateSelection} combines the deterministic baseline, accepted LLM, and validation-diversity lanes. The \texttt{LearningContract} copies statistical rules only from a frozen registry, never from free-form Planner output.\\[6pt]

\colorbox{red!20}{\textbf{\# Failure Semantics}}\\[2pt]
\begin{itemize}\setlength\itemsep{0pt}\setlength\topsep{2pt}
  \item \textbf{Stale task/revision/epoch:} reject the entire proposal.
  \item \textbf{Illegal action:} reject independently; legal siblings survive.
  \item \textbf{Timeout / invalid schema / no legal LLM action:} the LLM lane is empty; the deterministic baseline and diversity lanes remain active.
  \item \textbf{No lane has a legal action:} return control to the Manager for wait/close/stop/escalation.
\end{itemize}
\end{tcolorbox}

\subsection{Curator --- Post-Settlement Memory Writer}
\label{app:prompts_curator}

An asynchronous specialist invoked only after terminal evidence is committed. It runs in two modes: a per-settlement Strategy Memory Curator and a post-finalization Postmortem Curator.

\begin{tcolorbox}[promptbox=tag_colframe/tag_colback, title={Curator Prompt}, breakable]

\colorbox{green!20}{\textbf{\# Strategy Memory Curator}}\quad\textit{per-settlement}\\[2pt]
\textbf{Role.} You are Strategy Memory Curator. Convert committed terminal settlements into an append-only \texttt{StrategyBundleMemory} patch.\\
\textbf{Use} only the supplied settlement, frozen treatment identity, comparator, population, metric, assignment, and evidence-role references. Preserve both local diagnostic evidence and whole-tree causal evidence without conflating them; record negative, mixed, inconclusive, underpowered, unsafe, and non-promoted outcomes explicitly.\\
\textbf{Do not.} Alter raw evidence, infer current-task lift, create a control command, or block current experiment progress.\\
\textbf{Return.} Only the requested patch JSON.\\[6pt]

\colorbox{blue!20}{\textbf{\# Postmortem Curator}}\quad\textit{post-finalization}\\[2pt]
\textbf{Role.} You are Postmortem Curator. Run only after experiment finalization with a complete terminal trace, delivery result, failure/retry history, and user feedback.\\
\textbf{Propose} a \texttt{MethodologyMemoryPatch} that explains which operational method should be added, changed, or retained and cites the supporting trace. Distinguish a single-task incident from a reusable method; mark unresolved governance or approval questions.\\
\textbf{Do not.} Approve your own patch, modify a completed task, reinterpret settled statistics, or create runtime commands.\\
\textbf{Return.} Only the requested patch JSON.\\[6pt]

\colorbox{red!20}{\textbf{\# Governance Rules}}\\[2pt]
\begin{enumerate}\setlength\itemsep{0pt}\setlength\topsep{2pt}
  \item Accept only committed settlement references.
  \item Preserve positive, negative, mixed, inconclusive, diagnostic, and non-promoted tree-level evidence without collapsing them.
  \item Validate every patch with \texttt{scripts/patch\_validator.py} before append.
  \item Preserve provenance and conflicts; never rewrite raw evidence.
  \item A Curator failure must not block experiment progress.
  \item Only approved Methodology memory may fill unspecified future Runbook steps.
\end{enumerate}
\end{tcolorbox}

\section{Illustrative Memory Entries}
\label{app:memory_cases}

The following examples show the knowledge retained in Strategy Evidence and
Methodology Experience after curation. The values in the Strategy Evidence
entry are illustrative.

\begin{tcolorbox}[
    promptbox=interest_colframe/interest_colback,
    title={Strategy Evidence Entry},
    breakable
]
\textbf{Bundle comparison.}
Current ranking formula \texttt{S0} $\rightarrow$ enhanced ranking formula
\texttt{S1}. The comparison applies to the complete, versioned strategy
bundles rather than to an individual parameter or component.

\textbf{Whole-bucket evidence.}
Across the randomized experiment population, IPV lift was $0.3\%$, with a
$95\%$ confidence interval of $[-0.2\%, 0.8\%]$. Promotion required a
whole-bucket IPV lift of at least $1.0\%$, a positive confidence-interval lower
bound, and no guardrail violation. Although no guardrail was violated, the
observed lift was below the required threshold and the confidence interval
included zero. The complete \texttt{S1} bundle was therefore not promoted.

\textbf{Diagnostic evidence.}
A post-hoc analysis of users satisfying \texttt{pay\_cnt >= 3} showed an IPV
lift of $2.4\%$. Because this subgroup was not specified before randomization,
the result is retained as diagnostic evidence rather than as a validated causal
effect.

\textbf{Curated conclusion.}
\texttt{S1} is not validated as an improvement for the whole population. The
positive result for \texttt{pay\_cnt >= 3} remains an unresolved subgroup
signal and does not override the whole-bucket certificate. It may motivate a
targeted follow-up experiment in which the subgroup is specified before
randomization, but it cannot support a subgroup-level causal claim until such
an experiment is completed.
\end{tcolorbox}

\begin{tcolorbox}[
    promptbox=tag_colframe/tag_colback,
    title={Methodology Experience Entry},
    breakable
]
\textbf{Operation type.}
\texttt{observe}.

\textbf{Applicable scenario.}
The monitored primary or guardrail metric, such as transaction amount,
exhibits substantial day-to-day volatility, and the available observation
period is too short to support a stable conclusion.

\textbf{Observed issue.}
A positive or negative result from a single day may reflect ordinary daily
variation and may be reversed by subsequent observations. Treating such a
movement as a settled result can therefore lead to a premature conclusion or
an unnecessary intervention.

\textbf{Operational recommendation.}
Do not draw a conclusion from a single day's result. Continue observation
until the metric maintains the same direction for the predefined number of
consecutive days and satisfies the registered statistical decision rule. If
either condition remains unmet, keep the result unresolved and continue
observing within the approved task duration and resource budget. Directional
consistency guides the observation schedule but does not replace statistical
validation.

\textbf{Confidence state.}
\pill[ros]{draft}, pending confirmation from independent tasks. If consistent
outcomes recur across tasks, the entry may advance to
\pill[blu]{supported} and eventually to \pill[blu]{approved}. Until then, it
may be presented only as a hint and cannot override the current task's
statistical rules, budget, or approval requirements.
\end{tcolorbox}

\end{CJK*}
\end{document}